\documentclass{jkas}

\def\year{2025} 
\def\volume{---} 
\def\issue{---} 
\def\beginpage{1} 
\def\received{---} 
\def\accepted{---} 
\def\published{---} 
\date{Received \received; Accepted \accepted; Published \published}

\newcommand\ion[2]{{#1}\,{\sc #2}} 

\graphicspath{{./}{figures/}}

\usepackage{booktabs}

\title{%
K-DRIFT: Unveiling New Imagery of the Hidden Universe
}

\author[1,2$\star$]{Jongwan Ko}{0000-0002-9434-5936}
\author[1$\star$]{Woowon Byun}{0000-0002-7762-7712}
\author[1,2]{Kwang-Il Seon}{0000-0001-9561-8134}

\author[1,2]{Jihun Kim}{}
\author[1,2]{Yunjong Kim}{0000-0003-0009-5161}
\author[3,4,5]{Daewook Kim}{}
\author[6]{Seunghyuk Chang}{}
\author[7]{Dohoon Kim}{}
\author[8]{Il Kweon Moon}{}
\author[9]{Hyuksun Kwon}{}
\author[1,10]{Yeonsik Kim}{}
\author[1,2]{Kyohoon Ahn}{}
\author[1,2]{Gayoung Lee}{}
\author[1]{Yongseok Lee}{0000-0001-7594-8072}
\author[1]{Sangmin Lee}{}
\author[1]{Sang-Mok Cha}{0000-0002-7511-2950}
\author[1]{Dong-Jin Kim}{}
\author[1]{Kyusu Park}{}

\author[1,11]{Jaewon Yoo}{0000-0002-6841-8329}
\author[1]{Jae-Woo Kim}{0000-0002-1710-4442}
\author[1,2]{Jihye Shin}{0000-0001-5135-1693}
\author[1]{Sang-Hyun Chun}{0000-0002-6154-7558}
\author[1,10]{Yongmin Yoon}{0000-0003-0134-8968}
\author[1]{Jaehyun Lee}{0000-0002-6810-1778}
\author[1]{Kyungwon Chun}{0000-0001-9544-7021}
\author[1,12]{Jinsu Rhee}{0000-0002-0184-9589}
\author[1]{Sungryong Hong}{0000-0001-9991-8222}
\author[1]{Jongyeob Park}{}
\author[1]{Young-Beom Jeon}{}
\author[1]{Eon-Chang Sung}{}
\author[1,2]{Hong Soo Park}{0000-0002-3505-3036}

\author[1,2]{Seonwoo Kim}{}
\author[1,2]{GyeongGon Bahk}{}
\author[1,2]{Seri Yeon}{}

\affil[1]{Korea Astronomy and Space Science Institute, Daejeon 34055, Republic of Korea}
\affil[2]{Department of Astronomy and Space Science, University of Science and Technology, Daejeon 34113, Republic of Korea}
\affil[3]{Wyant College of Optical Sciences, University of Arizona, Tucson, AZ 85721, USA}
\affil[4]{Department of Astronomy and Steward Observatory, University of Arizona, Tucson, AZ 85721, USA}
\affil[5]{Large Binocular Telescope Observatory, Tucson, AZ 85721, USA}
\affil[6]{Center for Integrated Smart Sensors, Daejeon 34141, Republic of Korea}
\affil[7]{Green Optics Co., Ltd., Cheongju 28126, Republic of Korea}
\affil[8]{Center for Space Optics, Korea Research Institute of Standards and Science, Daejeon 34113, Republic of Korea}
\affil[9]{ADSOLUTION Co., Ltd., Daejeon 34168, Republic of Korea}
\affil[10]{Department of Astronomy and Atmospheric Sciences, College of Natural Sciences, Kyungpook National University, Daegu 41566, Republic of Korea}
\affil[11]{Quantum Universe Center, Korea Institute for Advanced Study, Seoul 02455, Republic of Korea}
\affil[12]{Institut d’Astrophysique de Paris, Sorbonne Université, CNRS, UMR 7095, 98 bis bd Arago, 75014 Paris, France}

\def\corrauthor{%
J. Ko, \email{jwko@kasi.re.kr}; W. Byun, \email{wbyeon@kasi.re.kr}
}

\def\runningauthor{%
Ko et al.
}

\def\runningtitle{%
K-DRIFT: Overview
}

\def\keywords{%
methods: observational --- techniques: image processing --- telescopes --- surveys --- galaxies: evolution 
}

\def\abstracttext{%
Low-surface-brightness (LSB) structures play a crucial role in understanding galaxy evolution by providing significant insights into galaxy interactions, the histories of mass assembly, and the distribution of dark matter. Nevertheless, their inherently faint nature, coupled with observational difficulties such as stray light interference and variations in the sky background, has significantly impeded comprehensive studies of LSB features. The KASI Deep Rolling Imaging Fast Telescope (K-DRIFT) project aims to address these observational challenges by developing off-axis freeform three-mirror telescopes and observational strategies specifically designed for LSB imaging surveys. The first generation of the K-DRIFT (K-DRIFT G1) has been successfully completed, and the forthcoming survey, scheduled to commence shortly, is expected to yield novel insights into the LSB universe. This paper outlines the scientific motivations of the project, discusses the technical challenges encountered, highlights the innovative solutions devised, and describes the future trajectory of the K-DRIFT.
}

\begin{document}
\jkashead 


\section{Introduction} \label{sec:intro}
\subsection{Motivation: The Importance of Investigating the Low-Surface-Brightness Universe} \label{sec:intro:motiv}
Within the context of the $\Lambda$ cold dark matter ($\Lambda$CDM) cosmological model, nearly all galaxies are observed to be enveloped by extensive and intricate low-surface-brightness (LSB) structures---remnants of past evolutionary processes such as mergers, accretions, and tidal interactions \cite[][]{1992ApJ...399L.117H,2007ApJ...666...20P,2008ApJ...689..936J,2013MNRAS.434.3348C,2014MNRAS.444..237P}. 
Numerous studies have documented the existence of stellar streams and filaments surrounding both the Milky Way \cite[e.g.,][]{2003ApJ...599.1082M,2007ApJ...658..337B,2007Natur.450.1020C} and the Andromeda galaxy \cite[e.g.,][]{2001Natur.412...49I,2002AJ....124.1452F,2007ApJ...671.1591I}. 
These structures, which emerge through the hierarchical assembly of galaxies, are expected to gradually disperse and coalesce into extensive stellar halos \cite[see][]{2008ApJ...689..936J}, offering a direct window into the long-term dynamical evolution of galaxies \citep[see][also references therein for a comprehensive review of tidal features]{2022MNRAS.513.1459M}.

Consequently, LSB structures are pivotal indicators of galaxy evolution, providing essential insights into galaxy interactions, mass assembly histories, and the underlying distributions of dark matter. 
However, their diffuse and extremely faint characteristics---typically several magnitudes fainter than the night sky brightness---pose significant challenges for detection using conventional imaging instruments and methodologies. 
For example, previous extensive imaging surveys, such as the Sloan Digital Sky Survey \cite[SDSS;][]{2000AJ....120.1579Y}, have exhibited a bias toward relatively luminous objects due to their limited detection capabilities ($\mu_r \lesssim 26.5$ mag arcsec$^{-2}$). Consequently, a substantial portion of the LSB universe remains inadequately explored, potentially resulting in an incomplete understanding of galaxy evolution. 

Investigating the LSB universe is crucial not only for enhancing our understanding of galaxy evolution but also for testing theoretical models of structure formation. 
The presence of LSB structures and their morphology provide valuable constraints that can be utilized to validate and refine cosmological simulations.
By examining these elusive features, we can bridge the gap between theoretical predictions and observational results.

\subsection{Challenges: The Difficulties of LSB Imaging } \label{sec:intro:challenge}
Valuable insights into the challenges of observing LSB features can be gained from pioneering work conducted nearly fifty years ago. Notably, \cite{1974AJ.....79..671K} and \cite{1984AJ.....89..919S} demonstrated impressive LSB imaging capabilities comparable to those of recent observations, successfully identifying stellar halos and dwarf galaxies. However, despite several decades of advancements in telescope technology, instrumentation, and data analysis methodologies, there has been no substantial improvement in the detection limit for LSB structures.
This is because the LSB imaging performance is not solely limited by photon statistics or spatial resolution; rather, they are fundamentally constrained by systematic uncertainties that hinder the detection of faint structures.

The principal technical challenges \cite[see][for a review of recent challenges in deep imaging]{2019arXiv190909456M} that lead to systematic errors in deep surface photometry, thereby impeding the detection and analysis of LSB structures, are outlined as follows:
\begin{itemize}
\item{\textbf{Contamination by stray light}: Unwanted photons from various sources, including the extended wings of the point spread function (PSF), multiple reflections between optical elements, scattered light within the telescope, complex diffraction patterns from obstructions in the telescope, and unshielded off-axis light, constitute fundamental systematic errors that restrict the ability to observe very low surface brightness.  
Conventional telescopes with obstructed beams are particularly vulnerable to this issue, which constitutes one of the most substantial challenges to be overcome in the quest for deep imaging.}
\item{\textbf{Flat-fielding residuals}: Inaccurate flat-fielding can introduce systematic errors during subsequent sky background estimation and subtraction. This problem worsens when the pixel response non-uniformity of the detector changes over time, especially if it is configured as an array with variations between chips. The resulting residuals in flat-fielding can mimic LSB objects, complicating the acquisition of reliable results in LSB studies. For example, flat-fielding errors need to be smaller than 0.01 $\%$ to reach a depth of $\sim$31 mag arcsec$^{-2}$, given a typical sky brightness in the r-band.}
\item{\textbf{Sky level variations}: The sky background results from a combination of many natural phenomena, including airglow lines in the upper atmosphere, zodiacal light, Galactic cirrus, scattered moonlight, and unresolved extragalactic background light. These factors, along with instrument-related issues, contribute to fluctuations in the sky background level, leading to inevitable spatial gradients across the field. This issue is particularly concerning in deep surface photometry of extended sources, as such gradients can cause excessive sky subtraction, resulting in the loss of LSB features in astronomical images.}
\end{itemize}
Addressing these challenges requires a comprehensive approach that encompasses LSB-optimized hardware design, effective observation strategies, and advanced data reduction techniques. 

\subsection{Strategies: Various Efforts for LSB Imaging}
Over the last decade, the field of deep imaging surveys has grown considerably to address challenges related to observational limitations. The study of the LSB universe has been propelled by several innovative projects, resulting in important discoveries and technological advancements. These efforts generally fall into two main categories: creating specialized instruments for LSB imaging and employing advanced data processing methods to improve the performance of existing telescopes. The first category includes dedicated LSB instrumentation such as:
\begin{itemize}
\item{The Dragonfly Telephoto Array \cite[][]{2014PASP..126...55A} is a novel observing system designed to achieve high-sensitivity, wide-field imaging utilizing multiple Canon 400 mm $f$/2.8 lenses. Its unique design significantly mitigates scattered light and internal reflections. By employing multiple small-aperture lenses, the system minimizes systematic errors and excels at detecting faint, diffuse structures. With a photometric depth reaching $\sim$29 mag arcsec$^{-2}$, it has enabled the discovery of very faint extended galaxies \cite[][]{2014ApJ...787L..37M,2015ApJ...798L..45V} and stellar halos \cite[][]{2016ApJ...830...62M}, leading to a variety of intriguing LSB studies.}
\item{The Burrell Schmidt Deep Virgo Survey \cite[][]{2017ApJ...834...16M} conducted deep imaging of the Virgo cluster, reaching a surface brightness limit of $\sim$29 mag arcsec$^{-2}$ in the $V$ band. The 0.6/0.9m Burrell Schmidt telescope is well suited for deep photometry on large scales, featuring a well-baffled closed-tube design, aggressive anti-reflection coatings on the optical elements of the camera, and a wide field-of-view (FoV) imager with a single CCD detector.}
\item{The Haloes and Environments of Nearby Galaxies (HERON) survey \cite[][]{2019MNRAS.490.1539R} examines LSB structures in the outskirts of galaxies within a volume of $\sim$50 Mpc, using the Jeanne Rich 0.7 m telescope. The instrument, designed with a fast f/3.2 beam and a wide 0.57 deg$^{2}$ FoV, also incorporates features to minimize scattered light. As a result, isophote fitting estimates suggest a limiting surface brightness level of $\sim$28--30 mag arcsec$^{-2}$ is achievable with exposure times of 1--2 hours.}
\item{The Los Alamos Low Surface Brightness Array \cite[][]{2020SPIE11447E..A2P} consists of a 16-element telephoto array, also utilizing Canon 400 mm $f$/2.8 lenses, similar to the Dragonfly Telephoto Array. A distinguishing feature is the incorporation of two sets of CCDs with differing pixel scales (7.25$^{\prime\prime}$ and 2.85$^{\prime\prime}$), which aids in alleviating source confusion in crowded regions while enhancing sensitivity.}
\end{itemize}
Deep imaging surveys, which utilize existing telescopes across a range of apertures (from sub-meter to 8-meter class), also contribute significantly to this field. Notable examples include:
\begin{itemize}
\item{The Stellar Tidal Streams Survey \cite[STSS; ][]{2019hsax.conf..146M} systematically detected and analyzed faint stellar streams to study galaxy mergers and dark matter distribution. Utilizing ten privately owned observatories equipped with telescopes ranging from 0.1- to 0.8-m in diameter, the survey achieved a surface brightness limit of $\sim$29 mag arcsec$^{-2}$ in the $r$ band with typical exposure times of 7--8 hours. This highlights that LSB observations are feasible even with modest-sized telescopes.}
\item{The Mass Assembly of Early-Type Galaxies with their Fine Structures (MATLAS) Survey \cite[][]{2015MNRAS.446..120D} is a deep optical imaging project that utilizes CFHT/MegaCam to investigate early-type galaxies beyond the Virgo Cluster. By leveraging its wide FoV, the survey employs large dithers to optimize the detection of extended LSB structures. While demonstrating the feasibility of LSB observations with a complex mosaic CCD camera, it also underscores the associated challenges in data processing and background estimation.}
\item{The Hyper Suprime-Cam (HSC) Deep and UltraDeep Survey \cite[][]{2018PASJ...70S...4A} was initially designed to investigate the evolution of galaxies, active galactic nuclei, and high-redshift supernovae. However, its combination of deep photometric depth and high-spatial resolution also renders it valuable for LSB studies at moderate distances. Notably, \cite{2023MNRAS.518.1195M} demonstrated that the UltraDeep layer, reaching $\sim$30.5 mag arcsec$^{-2}$, enables intra-group light studies beyond the local universe.}
\end{itemize}

\begin{figure}[t!]
\includegraphics[width=\linewidth]{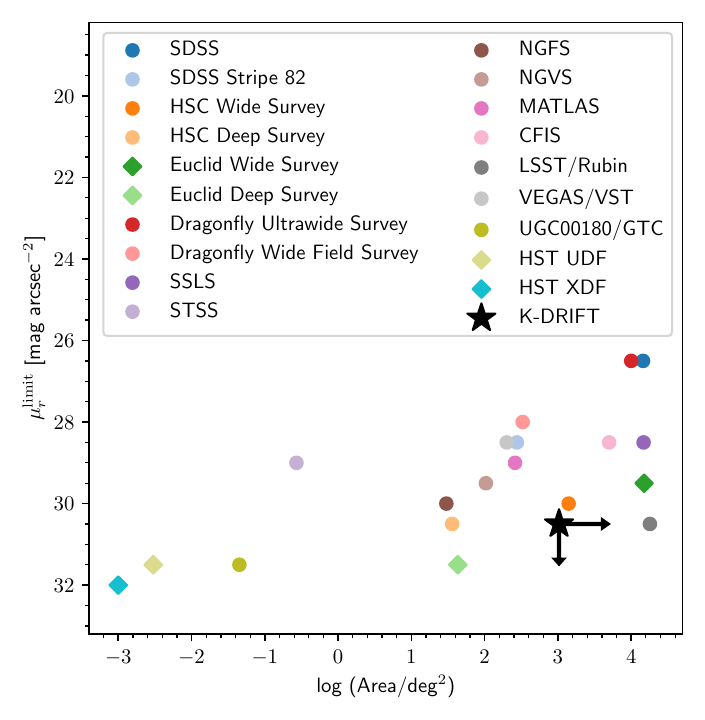}
\centering
\caption{Comparison of the 3$\sigma$ surface brightness limit at an angular scale of $10^{\prime\prime}\times10^{\prime\prime}$ as a function of the covered area for various imaging surveys. Surface brightness depths are estimated based on the $r$ band. The diamond shape represents space-based telescopes, while the circle shape represents ground-based telescopes.
The K-DRIFT survey presented in this paper is indicated by a black star, with its depth and coverage varying along the black lines depending on the specific observation strategy. This figure is a reproduced version of Figure 12 in \cite{2023A&A...671A.141M}. The imaging surveys shown in the figure include the following: SDSS \cite[][]{2000AJ....120.1579Y}, SDSS Stripe 82 \cite[][]{2014ApJS..213...12J,2016MNRAS.456.1359F}, HSC Wide Survey \cite[][]{2018ApJ...857..104G,2022PASJ...74..247A}, HSC Deep Survey \cite[][]{2022PASJ...74..247A}, Euclid Wide and Deep Survey \cite[][]{2022A&A...657A..92E}, Dragonfly Ultrawide Survey \cite[][]{2024ApJ...976...75S}, Dragonfly Wide Field Survey \cite[][]{2020ApJ...894..119D}, Stellar Stream Legacy Survey \cite[SSLS; ][]{2023A&A...671A.141M}, Stellar Tidal Streams Survey \cite[STSS;][]{2019hsax.conf..146M}, Next Generation Fornax Survey \cite[NGFS; ][]{2018ApJ...855..142E}, Next Generation Virgo Survey \cite[NGVS; ][]{2012ApJS..200....4F}, MATLAS \cite[][]{2015MNRAS.446..120D,2022A&A...662A.124S}, Canada-France Imaging Survey \cite[CFIS; ][]{2022A&A...662A.124S}, LSST/Rubin \cite[][10-yr integration]{2019ApJ...873..111I}, VEGAS/VST \cite[][]{2021arXiv210204950I}, UGC00180/GTC \cite[][]{2016ApJ...823..123T}, HST UDF \cite[][]{2006AJ....132.1729B}, HST XDF \cite[][]{2013ApJS..209....6I}.
\label{fig:f1}}
\end{figure} 

The 10-yr Legacy Survey of Space and Time \cite[LSST;][]{2019ApJ...873..111I}, which will be conducted at the Vera C. Rubin Observatory, and the Euclid space mission \cite[][]{2011arXiv1110.3193L,2022AAS...24030406B} are upcoming large-scale surveys poised to significantly enhance LSB studies through deep, wide-field imaging and unprecedented sensitivity. With their extensive sky coverage of nearly 20,000 deg$^2$, these surveys are expected to provide a robust statistical overview of the LSB universe.

Figure \ref{fig:f1} illustrates the sky coverage and surface brightness limits of various imaging surveys, including those introduced above.\footnote{See Appendix \ref{sec:app1} for details on the calculation of surface brightness limits at an angular scale of $10^{\prime\prime}\times10^{\prime\prime}$.} A trade-off often exists between achieving extensive sky coverage and obtaining ultra-deep photometric depth, as balancing these two objectives requires significant effort. Notably, missions such as LSST and Euclid are designed to excel in both areas and are expected to drive considerable progress in LSB research in the coming decades. 

Furthermore, these surveys differ in their primary scientific goals and technical specifications, including targeted photometric depth and FoV. More importantly, variations in data processing methodologies raise concerns regarding systematic consistency across different surveys \cite[][]{2019arXiv190909456M}. This issue is particularly relevant when employing optical systems with multiple lenses or conventional small reflective telescopes, where managing systematic uncertainties can be more intricate. In this context, initiatives focused on the development of new telescopes for systematic surveys, along with the improvement of observational techniques and data processing methods specifically designed for LSB imaging, are of significant importance.

\begin{figure*}[t!]
\centering
\includegraphics[width=\textwidth]{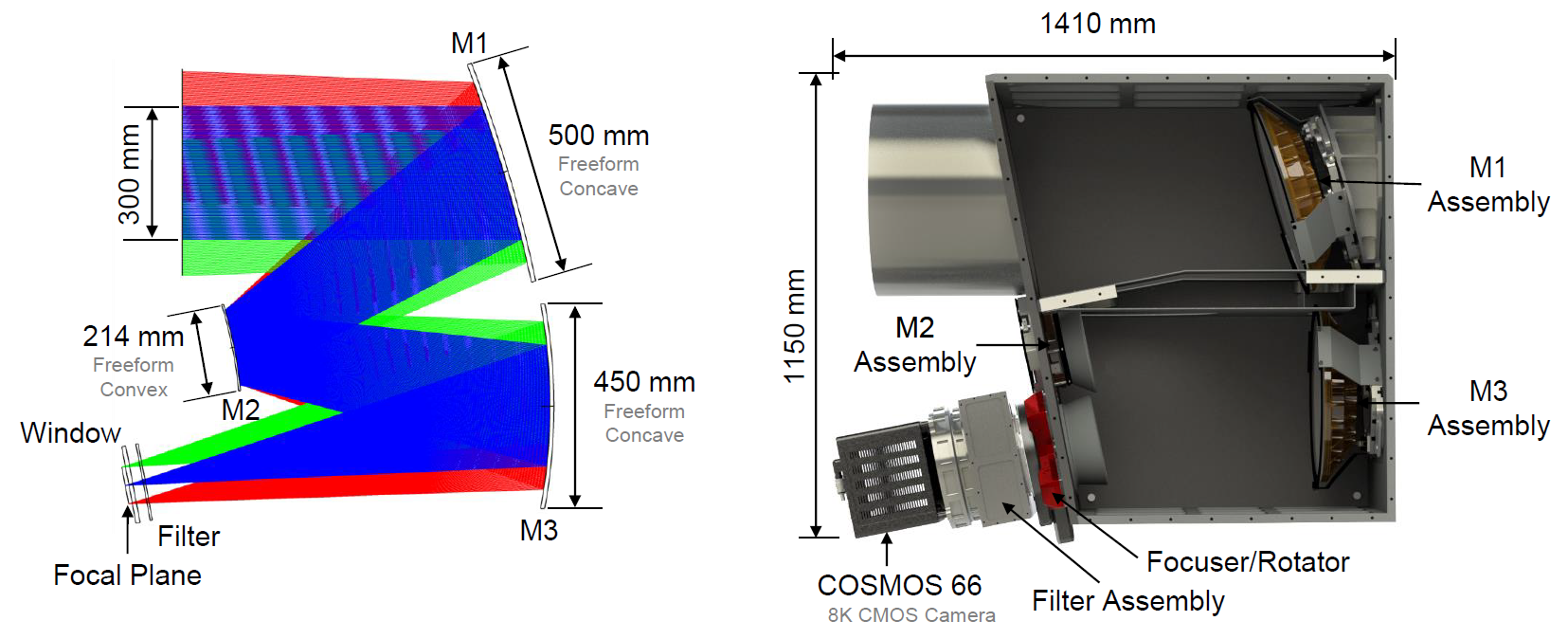}
\caption{Left: Optical design of K-DRIFT G1. Right: Opto-mechanical configuration with system parameters. The overall mechanical dimensions, including the camera, are $1410 \times 1150 \times 624$ mm. The diameters of the entrance pupil and the primary mirror are 300 mm and 500 mm, respectively.
\label{fig:f2}}
\end{figure*}

\section{K-DRIFT Framework: Advancing Optimal LSB Observations} \label{sec:frame}
Leveraging insights from previous research, the Korea Astronomy and Space Science Institute (KASI) initiated the KASI Deep Rolling Imaging Fast Telescope (K-DRIFT) project in 2019. This project presents a novel observational framework aimed at addressing key challenges in LSB imaging. The strategy is based on the following three core concepts:
\begin{itemize}
\item{\textbf{Wide-field imaging}: A large FoV allows for the simultaneous capture of extensive LSB structures while ensuring the availability of blank sky regions for accurate sky estimation and subtraction. Additionally, fast optics with a small focal ratio and large pixel scale enhance the ability to image large FoV, collecting more photons per pixel and thereby enabling high sensitivity to surface brightness even with small apertures.}
\item{\textbf{High-contrast imaging}: An off-axis reflective optical system significantly reduces internal reflections and scattering by eliminating diffraction artifacts caused by the central obstruction from the secondary mirror and its supports. This unobstructed optical path simplifies the PSF by reducing energy in the PSF's wings, enhancing contrast for faint features. Additionally, well-designed baffles and anti-reflection coatings on the telescope's optical elements further reduce stray light interference.}
\item{\textbf{Flat imaging}: Accurate flat-field correction is crucial for detecting and analyzing LSB features, as it corrects for variations in pixel sensitivity across imaging sensor. Typically, three types of master flats are used for this correction. Among these, the dark-sky flat is considered the most precise compared to twilight or dome flats, as it better represents the actual observation conditions, thereby improving accuracy \citep{2016ApJ...823..123T, 2019A&A...621A.133B}. Dark-sky flats are generated by combining the sky background from scientific images after a careful masking process. However, the night sky may exhibit complex local variations, which can still introduce photometric inaccuracies when using dark-sky flats for calibration. Therefore, an optimized observation strategy is necessary to obtain accurate dark-sky flats. This approach can help eliminate camera artifacts that vary with exposure time. Reducing flat-fielding errors to below 0.1 $\%$ leads to more reliable estimates and subtractions of the sky background.}
\end{itemize}

These ideas emphasize the goals behind the development and operation of the K-DRIFT telescope, which seeks to address the difficulties related to LSB imaging. The initial prototype (K-DRIFT Pathfinder) was tested through on-sky observations at the Bohyunsan Optical Astronomy Observatory (BOAO) from 2021 to 2022, achieving a surface brightness limit of $\sim$28.5 mag arcsec$^{-2}$ with two hours of integration time \cite[see][]{2022PASP..134h4101B}. 
The main model, `K-DRIFT Generation 1 (G1)', has demonstrated considerable improvements over the prototype, including a FoV that is 20 times larger. We have achieved a significant milestone by successfully manufacturing, assembling, and aligning all optical components of K-DRIFT G1. In order to perform the southern sky survey, two G1 units will be transported to Chile in the 2nd half of 2025.
The following sections provide details about the instruments and the survey strategy.

\begin{table}[htb]
  \centering
  \caption{K-DRIFT G1 specifications}
  \label{tab:optical-parameters}
  \begin{tabular}{ll}
    \toprule
    Parameter & Measurement \\
    \midrule
    Entrance pupil / Aperture diameter 
      & 300 / 500\,mm \\
    Focal length 
      & 1050\,mm \\
    Focal ratio 
      & 3.5 \\
    Field of view 
      & $4.43^\circ\mathrm{(H)} \times 4.43^\circ\mathrm{(V)}$ \\
    Image area 
      & $81.2\,\mathrm{mm} \times 81.2\,\mathrm{mm}$ \\
    Pixel scale 
      & $1.96'' / 10\,\mu\mathrm{m}$ \\
    \bottomrule
  \end{tabular}
\label{tab:t1}
\end{table}

\subsection{Instrument Outline} \label{sec:frame:inst}
\subsubsection{Optical Design} \label{sec:frame:inst:optics}
To effectively identify LSB structures, the K-DRIFT telescopes employ a linear-astigmatism-free three-mirror system \cite[LAF-TMS; ][]{10.1117/12.2023433,2020PASP..132d4504P}, which features a 300 mm aperture and an off-axis freeform three-mirror design---consisting of a concave primary mirror (M1), a convex secondary mirror (M2), and a concave tertiary mirror (M3). This unobstructed configuration, along with the freeform mirrors, minimizes stray light and reduces the wings of the PSF in the image plane, while also providing a wide FoV with consistent PSFs across the entire FoV. 

The diffraction limit for a 300 mm telescope operating at a wavelength of 550 nm is roughly 0.46 arcseconds, which represents the best possible seeing conditions at ground level. This indicates that the spatial resolution on the ground remains consistent, regardless of whether the aperture is 0.3-m or 10-m, since they are all constrained by atmospheric conditions, unless adaptive optics are employed. For extended structures, like LSB objects in the nearby universe, the imaging speed is influenced not only by the aperture size but also by the focal ratio. Furthermore, for telescopes intended for wide field surveys, the overall efficiency is a result of the interplay between the aperture, focal ratio, and the required resolution. Ultimately, we defined the key parameters of the optical design while considering the difficulties involved in fabricating a freeform optical system.

\begin{figure}[t!]
\includegraphics[width=\linewidth]{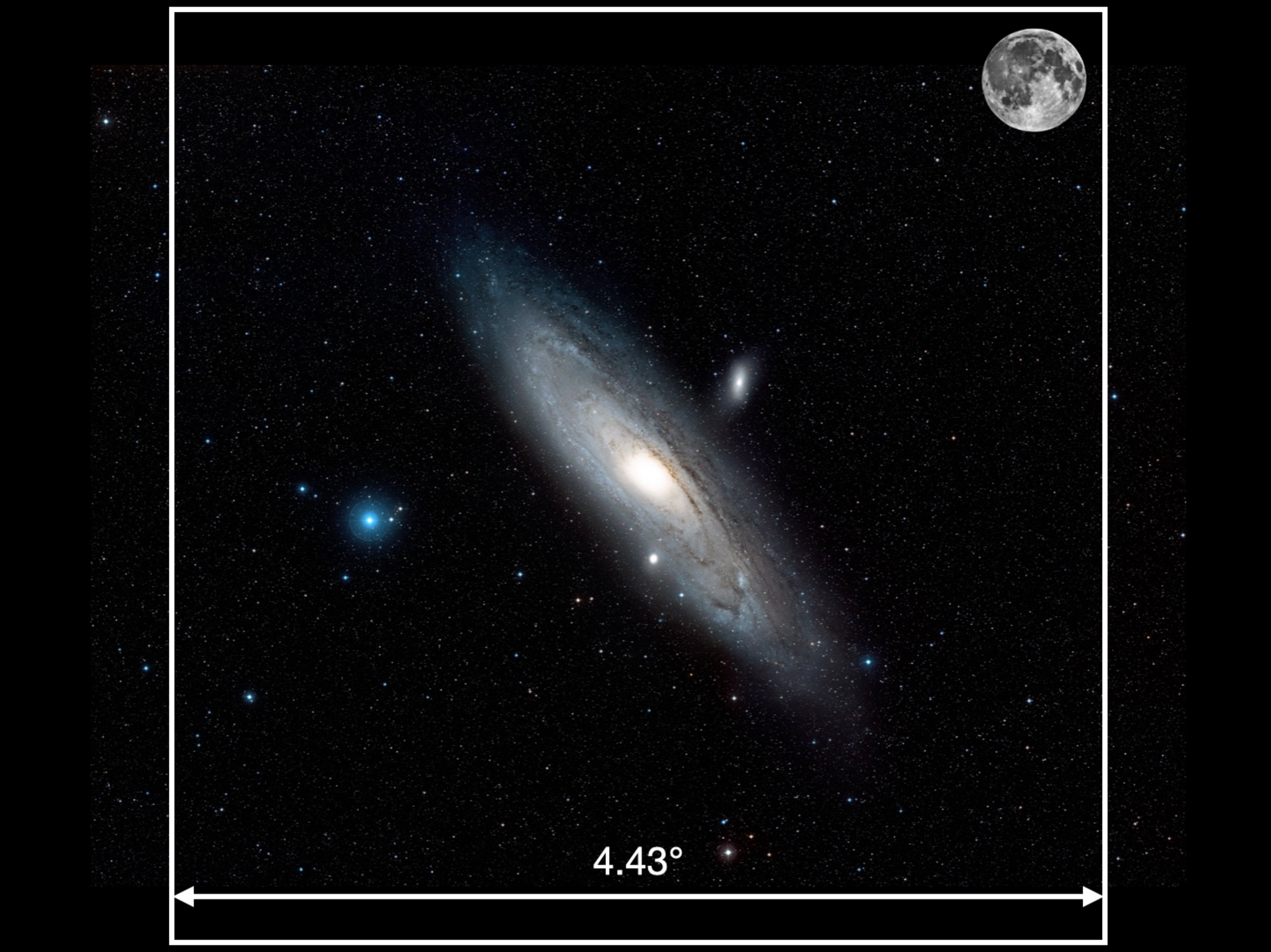}
\centering
\caption{Comparison of the FoV of K-DRIFT G1 (solid white box) with the Andromeda galaxy and the full moon. K-DRIFT G1 can capture nearly 100 times the area of the full moon in a single exposure, enabling efficient mapping of extended LSB structures. Credit: ESA/Hubble \& Digitized Sky Survey 2. Acknowledgment: Davide De Martin (ESA/Hubble)
\label{fig:f3}}
\end{figure}

Figure \ref{fig:f2} illustrates the setup of K-DRIFT G1, with specific details provided in Table \ref{tab:t1}. The entrance pupil has a diameter of 300 mm, while the primary mirror features a clear aperture of 500 mm. The focal ratio is $f$/3.5, and it is combined with a CMOS sensor of 81.2 mm $\times$ 81.2 mm at the focal plane, producing a wide FoV of $4.43^\circ\times4.43^\circ$ (see Figure \ref{fig:f3}). The optical design achieves a spot diameter, characterized by the diffraction limit at the focal plane, ranging from 2.7 to 8.9 $\mu$m, with an average of 6.0 $\mu$m, which is significantly smaller than the pixel size of 10 $\mu$m. To minimize stray light, baffles are strategically placed in front of the entrance, in the middle of the M1-to-M2 optical path, and in front of the detector. The total weight of the opto-mechanical structure is approximately 285 kg, and all components have been confirmed to remain stable under the observation environment. Additionally, we conducted an analysis of the telescope's optical performance in relation to temperature variations in the observing environment and changes in the telescope's orientation through finite element analysis. The results showed that these factors did not significantly impact the PSF (Y. Kim et al. 2025, in preparation). 

\subsubsection{Camera} \label{sec:frame:inst:camera}
The Teledyne COSMOS-66 is a complementary metal-oxide-semiconductor (CMOS) camera specifically designed for high-performance ground-based astronomical imaging applications \cite[][]{2022SPIE12191E..0IC}. It is equipped with a large-format sensor with a resolution of $8120\times8120$ pixels and a 10 $\mu$m pixel pitch, resulting in a pixel scale of 1.96$^{\prime\prime}$. The sensor is a thinned, back-side illuminated device featuring dual conversion gain, which facilitates low read noise, high sensitivity, and the capability to operate in high dynamic range modes. Additionally, the camera incorporates a fused silica window that is treated with a broadband-optimized anti-reflective coating.

Teledyne's specifications indicate a read noise of $\leq$1 $e^-$ and a dark current of $\leq$0.05 $e^-$/pix/s at $-25^\circ\mathrm{C}$. However, recent performance measurements have demonstrated an effective read noise of 2.9 $e^-$ and a dark current of $\leq$0.12 $e^-$/pix/s at the same temperature \cite[][]{Layden+25}. The camera ensures high quantum efficiency (QE) of over 50\% across the 250--800 nm range, peaking at 90\% around 600 nm. 
With low noise performance similar to large-format charge-coupled devices (CCDs), high frame rate, and large pixels, the COSMOS-66 camera is particularly well-suited to large-scale LSB surveys.

\subsubsection{Filter Sets} \label{sec:frame:inst:filters}
\begin{figure}[t!]
\includegraphics[width=\linewidth]{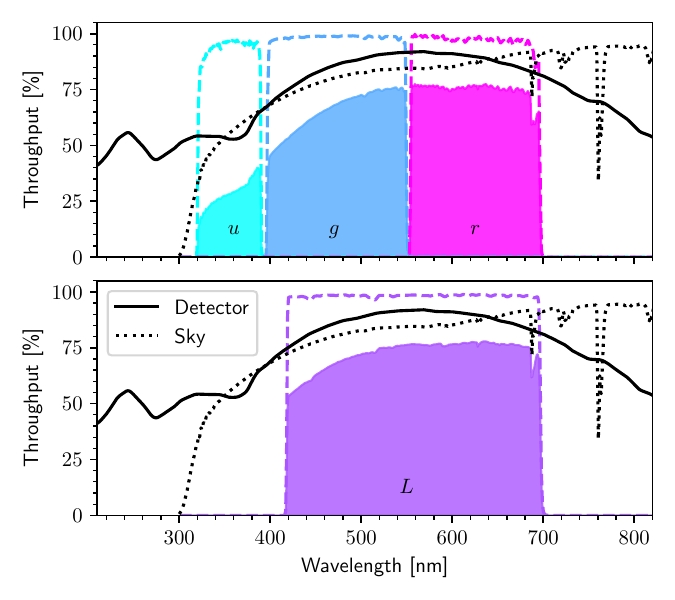}
\centering
\caption{System throughput as a function of wavelength, showing contributions from the filter transmission (dashed color-coded lines), detector QE (solid black lines), and atmospheric transmission at 1.2 airmass (dotted black lines).
\label{fig:f4}}
\end{figure}

The filter sets currently available for the K-DRIFT G1 system comprise Sloan $ugr$ filters and a Luminance ($L$) filter, all custom-produced by Chroma Technology Corp. Each filter has dimensions of $115\times115\times5$ mm. The $L$ filter's broad bandpass makes it particularly useful for detecting LSB signals. Figure \ref{fig:f4} shows the throughput of these filters, taking into account the detector's QE and atmospheric transmission. 

K-DRIFT G1 is equipped with a motorized filter slide, enabling each telescope to accommodate up to two filters. The initial filter pairings will be $L\&g$ and $L\&r$. Under dark conditions, both telescopes will perform observations using the $L$ filter to enhance LSB detection, while in brighter conditions, the $g$ and $r$ filters may be utilized for monitoring variable objects. 
In the future, we plan to add narrow-band filters, such as H$\alpha$ and [\ion{O}{iii}], which will replace the $L$ filters for observations of the ionized interstellar and circumgalactic medium. These narrow-band observations are intended to survey ionized gas at redshifts below 0.01. For instance, the H$\alpha$ survey will be conducted using two narrow-band filters: one centered at 6596\text{\AA} for the emission line and the other at $\sim$6696 (or 6496)\text{\AA} for the continuum, each with a width of $\sim$100\text{\AA}. Alternatively, an $r$-band filter can be used for the continuum subtraction. Follow-up observations may be carried out for detailed studies of specific targets identified in the survey, either by employing narrower-band filters with a width of $\sim$50\text{\AA} on K-DRIFT.

\subsubsection{Focuser} \label{sec:frame:inst:focus}
The focuser is a custom-designed ESATTO unit produced by PrimaLuceLab. It can support a maximum payload of 40 kg and operates effectively within a temperature range of $-20^\circ\mathrm{C}$ to $60^\circ\mathrm{C}$. The unit also features a motorized rotator that allows the attached filter and camera to rotate at speeds of up to 1$^\circ$ per second. 

As described in the following section, K-DRIFT G1 employs a unique observation method that includes camera rotation. If artifacts are introduced from the filter, and the filter and the camera do not move together, their positions and shapes will vary between exposures, complicating accurate corrections. This could degrade the quality of individual frames and significantly impact LSB studies. To prevent this issue, we have implemented a design that allows the filter and camera to rotate together, ensuring consistent image quality.

\subsubsection{Mount} \label{sec:frame:inst:mount}
\begin{figure}[t!]
\includegraphics[width=\linewidth]{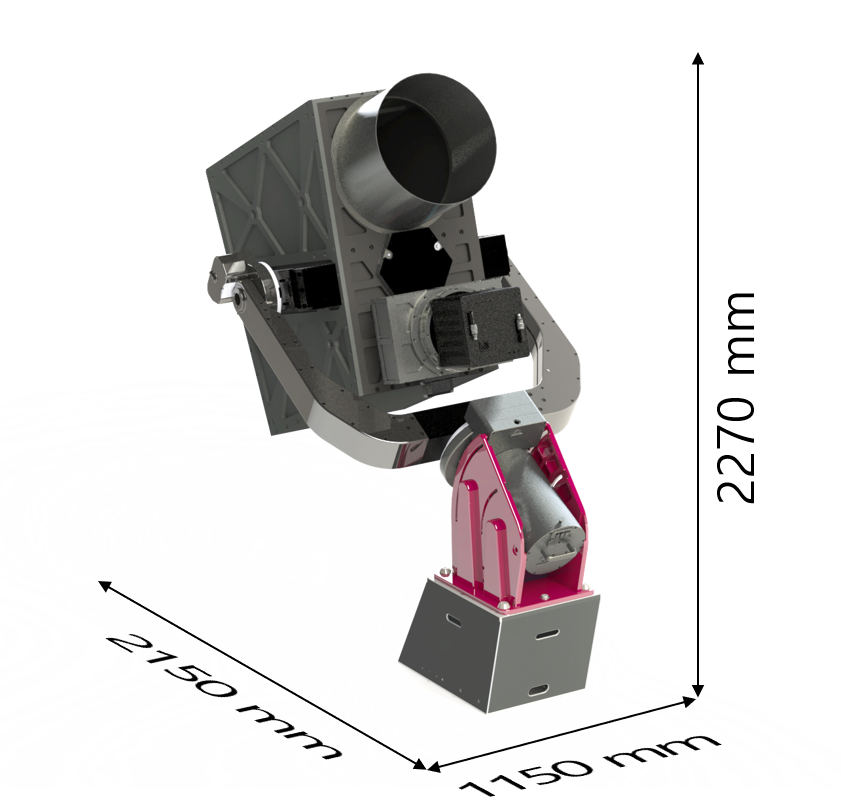}
\centering
\caption{Telescope rendering of the K-DRIFT G1. Two identical telescopes will conduct the initial LSB imaging survey at El Sauce Observatory in Chile.
\label{fig:f5}}
\end{figure}

The telescope will be mounted on a Paramount Taurus 700 model (see Figure \ref{fig:f5}), produced by Software Bisque Inc. This mount features a direct-drive equatorial fork design and is one of the largest available commercially. It can achieve a slew speed of up to 30$^\circ$ per second, with an expected pointing accuracy of 20 arcseconds or better. To enhance weight distribution and minimize distortion during operation, the telescope bodies are secured to the forks using a custom-designed mounting plate. Each mount operates independently, controlled by its own mount control PC.

\subsection{Survey Strategy} \label{sec:frame:survey}
\subsubsection{Site} \label{sec:frame:survey:site}
The K-DRIFT G1 telescopes will be installed at El Sauce Observatory in the Coquimbo Region of Chile, which is operated by Obstech.\footnote{\url{https://obstech.cl}} The exact location and typical weather conditions are as follows:
\begin{itemize}
\item{Latitude $-30^\circ\ 28^\prime\ 16^{\prime\prime}$}
\item{Longitude $-70^\circ\ 45^\prime\ 47^{\prime\prime}$}
\item{Elevation 1600 m}
\item{Wind speed during nighttime $<$5 m/s, mostly $\sim$0.5 m/s}
\item{Humidity during nighttime $\sim$10--70\%}
\item{Temperature during nighttime $\sim$0--20$^\circ\mathrm{C}$}
\item{Night sky brightness $\sim$21.5--22 mag arcsec$^{-2}$}
\item{Seeing $\sim$1.3$^{\prime\prime}$}
\item{Available nights per year $>$300 nights}
\end{itemize}

This site is well-suited for conducting LSB imaging surveys, as it provides access to the darkest night skies and allows for extended observation times. 
Although the seeing at this location is slightly poorer than the average for other sites in Chile, it will not pose a significant limitation to our LSB research.
In fact, K-DRIFT G1 has a pixel scale of approximately 2$^{\prime\prime}$, which is larger than the typical seeing size. This indicates that K-DRIFT G1 is limited by its pixel resolution, as it balances angular resolution with a wide FoV of about 20 deg$^{2}$. This trade-off reflects that our scientific focus is primarily on LSB features in the nearby universe. For example, to capture a low surface brightness profile (e.g., as faint as 30 mag arcsec$^{-2}$) of a target galaxy in a single image rather than through a mosaic observation, it is essential to cover an area ranging from several arcminutes to several degrees.
The following sections describe the optimized operational strategy aimed at enhancing data quality at this site.

\subsubsection{Observation Method: Rolling Dithering} \label{sec:frame:survey:dithering}
As mentioned earlier, minimizing systematic uncertainties in observations is crucial for advancing LSB studies. Among the three key concepts of K-DRIFT, introduced at the beginning of Section \ref{sec:frame}, the first two were addressed through our innovative optics design. To achieve the remaining third---flat imaging---new approaches beyond traditional observation and data processing methods are required. 

\begin{figure}[t!]
\includegraphics[width=\linewidth]{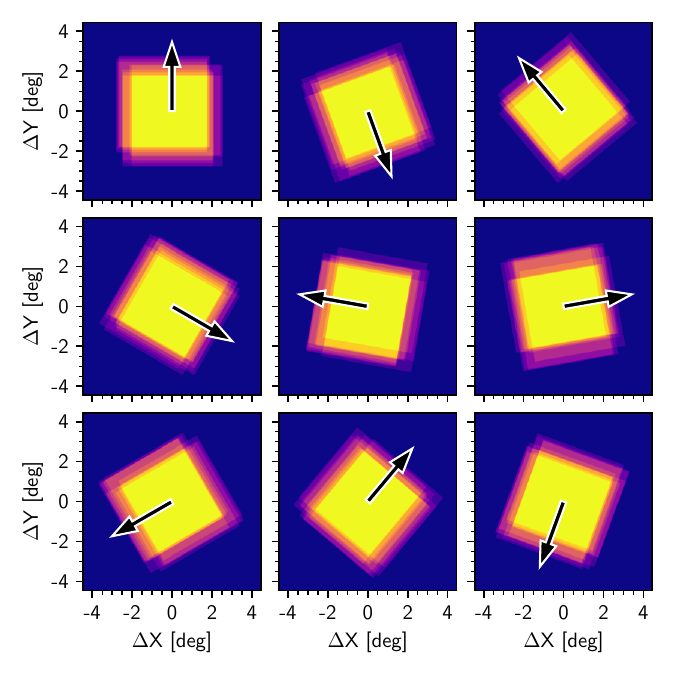}
\centering
\caption{Schematic example of the rolling dithering method. The observation sequence proceeds from top left to bottom right, with arrows indicating the camera's position angle, which rotates by 160$^\circ$ in sequence. Each panel consists of ten frames with random $xy$-shifts of up to 40 arcminutes.
\label{fig:f6}}
\end{figure} 

To significantly improve the accuracy of flat field correction, we developed a new observational technique called ``rolling dithering''. This method incorporates camera rotation into conventional dithering procedures. A brief comparison of the workflows is presented below: 
\begin{itemize}
\item{Conventional dithering sequence: exposure $\rightarrow$ $xy$-shift $\rightarrow$ exposure $\rightarrow$ $xy$-shift $\rightarrow$ ...}
\item{Rolling dithering sequence: (exposure $\rightarrow$ $xy$-shift) $\times$ dozens of times $\rightarrow$ camera rotation $\rightarrow$ (exposure $\rightarrow$ $xy$-shift) $\times$ dozens of times $\rightarrow$ camera rotation $\rightarrow$ ...}
\end{itemize}

Figure \ref{fig:f6} illustrates how the coverage and the camera's position angle change during a rolling dithering sequence. The default rotation angle is 160$^\circ$, and a complete observation sequence consists of nine different position angles. 
The rotation angle and xy-shift size can be adjusted based on scientific objectives and observational conditions, including the target, survey cadence, and sky background level. In our baseline approach, we will implement random xy-shifts within a $40^\prime \times 40^\prime$ region centered on the reference coordinate. This region is several times larger than the anticipated angular diameter of a Milky Way-sized galaxy at a distance of 20 Mpc. This configuration is designed to minimize source overlap across individual images, thereby facilitating the generation of high-quality dark-sky flats free from unwanted pixels and artifacts.

This method reduces the fully stacked, overlapped area by approximately 20\% compared to conventional dithering, as the outer part of the circular pattern receives significantly less exposure time. However, by rotating the camera, fluctuations in the sky background can be effectively averaged out, providing a more accurate flat frame and reducing flat-fielding errors by nearly a factor of five. A detailed validation and discussion of this method can be found in \cite{2025PASP..137e4502B}. 

In addition to reducing flat-fielding error, rolling dithering offers two additional advantages: First, the final coadded image will be azimuthally homogeneous, mitigating sky subtraction errors in individual images and improving subsequent analyses, such as isophote fitting for large galaxies. Second, this technique, which combines random $xy$-shifts in arcsec (or sub-pixel) scale with rotation of the pixel grid, potentially enables variable-pixel linear reconstruction, also known as ``Drizzle''. This capability is especially advantageous for K-DRIFT G1, which has a relatively large pixel scale and consequently undersamples the PSF. 

\subsubsection{Operation Plan} \label{sec:strategy:operation}
The operational plan is categorized into two main types of observations: bright-time and dark-time. Bright-time observations mainly involve monitoring supernovae and variable photometry, which are less affected by systematic errors and therefore allow for more flexible observation scheduling. For instance, the observation cadence can vary depending on the target, the airmass limit can be relaxed, and rolling dithering may not be necessary. Conversely, dark-time observations, which focus on LSB studies, require a more structured approach. These observations must be conducted under optimal sky conditions to minimize systematic uncertainties. This section outlines the strategies used to ensure high-quality LSB observations.

\begin{figure}[t!]
\includegraphics[width=\linewidth]{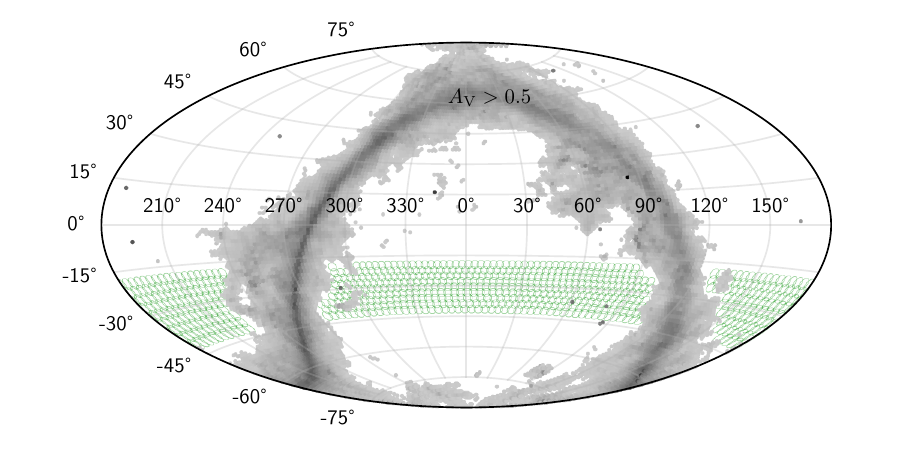}
\includegraphics[width=\linewidth]{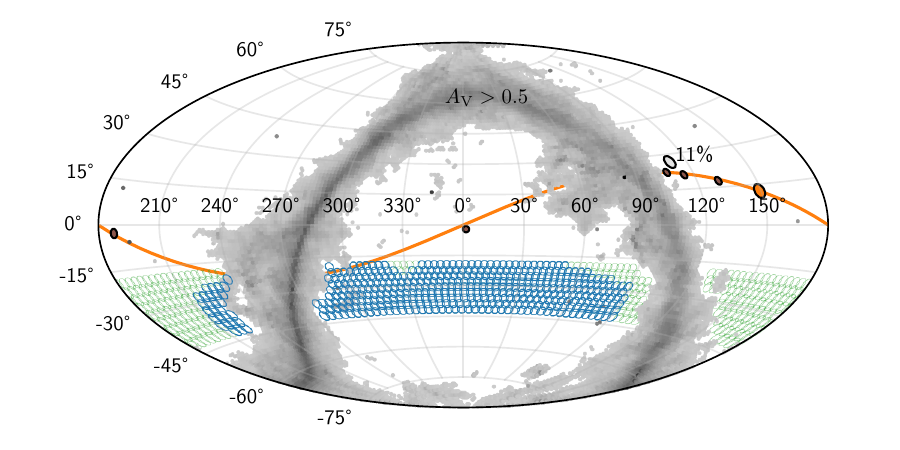}
\includegraphics[width=\linewidth]{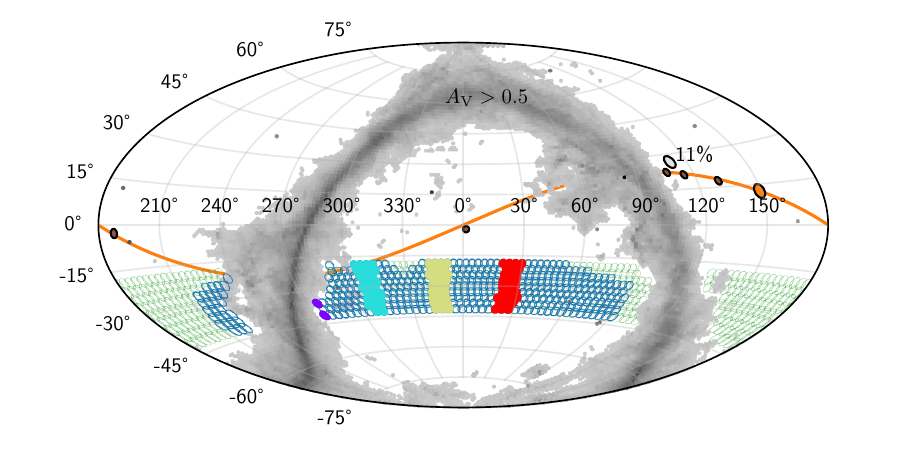}
\centering
\caption{Field selection process illustrated using an example from 2025-08-19 UTC. Top: annually observable fields (green circles), defined as regions within the declination range of $-20^\circ$ to $-40^\circ$ while avoiding Galactic cirrus (grayscale). Middle: daily observable fields (blue circles), determined by accounting for the influence of the sun (orange circle), moon (gray-shaded circle; with illumination noted), and planets (brown circles) along the ecliptic plane (solid orange line). Bottom: four groups of fields that can be scheduled for observation (color-coded filled circles), selected based on their culmination time during the night. 
\label{fig:f7}}
\end{figure} 

To facilitate a well-organized survey over nearly five years, we created a simulation for field (or target) selection. The selection criteria include constraints on annual, daily, and hourly timescales. To maximize the altitude of the field during observations, we limited the declination range to between $-20^\circ$ and $-40^\circ$, allowing us to observe targets closer to the zenith and thereby reduce airmass and airglow interference. Additionally, to avoid the detrimental effects of Galactic cirrus on LSB studies, we excluded fields with $A_V > 0.5$ mag, resulting in an annually observable area of approximately 5000 deg$^2$.

Daily observations should be conducted in dark conditions, which are defined as when the sun is at least $18^\circ$ below the horizon. The moon's impact is assessed based on its phase: if its illumination is below 20\% (darkest), it can be ignored; for 20-50\% (dark) illumination, fields within 50$^\circ$ of the moon should be avoided; and for illumination above 50\%, observations should only be conducted when the moon is below the horizon. In fact, observations at shorter wavelengths (B- to R-band) during dark and gray times (50-80\% illumination) are minimally affected by the moon if it is more than 50-60$^\circ$ away.\footnote{See https://www.eso.org/sci/observing/phase2/ObsConditions.html} 
To minimize the effect of zodiacal light on sky brightness, we also exclude fields where the surface brightness of zodiacal light exceeds $\sim$22 mag arcsec$^{-2}$. After applying these criteria, the remaining observable fields are further refined to ensure they reach an altitude of at least 50$^\circ$ during the night.

\begin{figure}[t!]
\includegraphics[width=\linewidth]{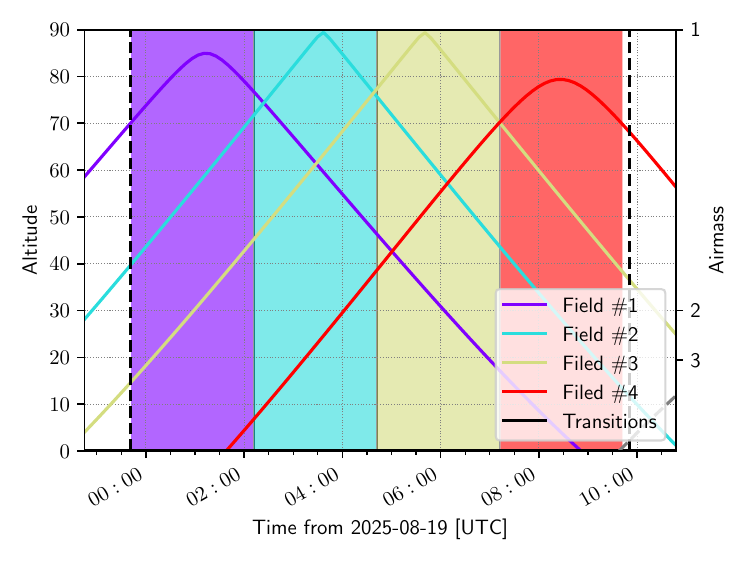}
\centering
\caption{Example of the visibility of the four scheduled fields on 2025-08-19 UTC. Each field reaches culmination during its scheduled observation session, which lasts approximately 2.5 hours, including overhead time. The dashed vertical lines represent the astronomical twilight times.
\label{fig:f8}}
\end{figure}

The standard observation duration for each field per night is about 2.5 hours, including overhead time. With an average daily observability of around 6.5-9.5 hours at the site, up to four fields can be observed each night. Each field is scheduled to culminate for optimal imaging conditions, with the first field selected to culminate in about an hour and subsequent fields scheduled at approximately 2.5-hour intervals.

Figure \ref{fig:f7} illustrates an example of the field selection process for the night of 2025-08-19 UTC. The top panel shows the annually observable area divided into circular subregions—each with a radius of approximately $\sim$2$^\circ$—representing the final coverage with the deepest and most consistent photometric depth achieved through rolling dithering, spaced roughly 3$^\circ$ apart.
The middle panel includes the ecliptic plane and solar system objects, highlighting fields that avoid the influence of the sun and moon while minimizing zodiacal light contamination. A gap near $\mathrm{R.A.} = -30^\circ$ indicates fields excluded due to Gegenschein contamination. The bottom panel highlights the four groups of fields that will sequentially pass through culmination during the observation. One field selected in each group is scheduled for observation, with priority based on proximity to $\mathrm{Decl.} = -30^\circ$. 

Figure \ref{fig:f8} presents the visibility of the four scheduled fields, using the same color scheme as the bottom panel of Figure \ref{fig:f7}. All scheduled fields maintain an altitude above 70$^\circ$ throughout their observation periods. On this night, the final observation session ends just before morning twilight. If there is a significant gap before twilight, we implement two policies: if the remaining time is less than 30 minutes, dark-time observations end early and either calibration observations are interleaved or pre-selected monitoring field observations are performed; otherwise, observations continues as long as the last scheduled field stays above 60$^\circ$.

\begin{figure}[t!]
\includegraphics[width=\linewidth]{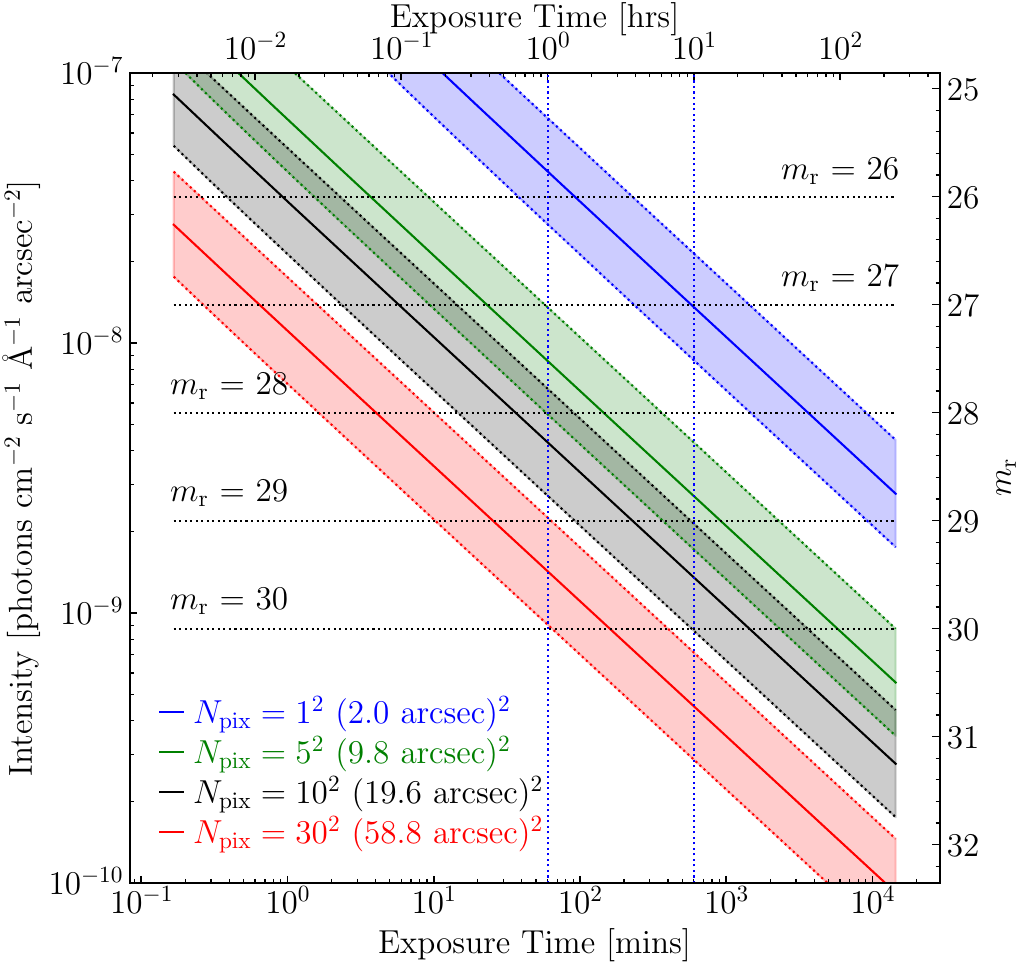}
\centering
\caption{SDSS $r$-band 3$\sigma$ sensitivity as a function of integration time for different angular scales: 1$\times$1 pixels (corresponding to $\sim$$2^{\prime\prime}\times 2^{\prime\prime}$), 5$\times$5 pixels ($\sim$$10^{\prime\prime}\times10^{\prime\prime}$), 10$\times$10 pixels ($\sim$$20^{\prime\prime}\times20^{\prime\prime}$), and 30$\times$30 pixels ($\sim$$59^{\prime\prime}\times59^{\prime\prime}$). The shaded region represents the range calculated when the sky background value varies by $\pm$1 mag arcsec$^{-2}$.
\label{fig:f9}}
\end{figure}

In Figure \ref{fig:f9} we present sensitivity calculations based on the K-DRIFT G1 specifications and the standard sky conditions at the observing site. To reach a surface brightness limit of $\geq$ 29 mag arcsec$^{-2}$ in the $r$ band, approximately ten hours of total integration time per field is required. These calculations were further validated by the results of \cite{2022PASP..134h4101B}, which report a surface brightness limit (3$\sigma$, $10^{\prime\prime}\times10^{\prime\prime}$) of approximately 27.3 mag arcsec$^{-2}$ after two hours of integration at BOAO. The sky background at BOAO is at least 1.5 magnitudes brighter than that at El Sauce Observatory in Chile, leading to an estimated imaging depth of 29--30 mag arcsec$^{-2}$ for K-DRIFT G1.

Given the extensive time required to survey the entire 5000 deg$^2$ with ten hours of integration per field, we decided to prioritize early science goals by initially targeting approximately 100 pre-selected fields. These fields were selected to achieve the target surface brightness depth of 29-30 mag arcsec$^{-2}$ in the $r$-band during the first year. The initial scientific programs will focus on ultra-diffuse galaxies (UDGs) in different environments (including the Local Void), nearby massive galaxies, and galaxy clusters. 
The scientific objectives of K-DRIFT G1 are detailed in the subsequent section.
 
Once the early science programs for the 100 selected fields are completed, we will proceed to the main survey. This extensive survey is designed to increase the sample size rather than to explore the existing data in greater detail, extending the coverage beyond the initially chosen fields without further improving the surface brightness limit. However, there may be strategies available to achieve greater surface brightness depth for specific fields. Figure \ref{fig:f1} illustrates the variations in depth and survey coverage for each approach. At the end of the initial survey, we will decide on the strategy for expanding the survey area and exploring deeper surface brightness levels.

\section{K-DRIFT Science Themes} \label{sec:science}
The K-DRIFT project aims to thoroughly explore the LSB universe to address fundamental questions across various areas of astrophysics. 
Further details on these scientific objectives will be described in forthcoming papers of this volume. This section, however, provides a brief overview of the primary scientific objectives.
\subsection{Galaxy Formation and Evolution}
Analyzing faint tidal features and remnants of mass accretion around galaxies is vital for understanding their mass assembly history. Tidal structures like stellar streams, tails, and shells provide insights into past mergers, including their mass ratios and orbital paths. Additionally, stellar halos, which are formed from the accumulation of these structures, retain information about even older merger events. The role of mergers in galaxy evolution is further clarified through comparative studies that examine how recent merger events relate to various physical properties of galaxies. Extending this approach to larger scales, analyzing the properties of intra-cluster light (ICL) within galaxy clusters allows us to deduce the evolutionary history of the brightest cluster galaxies and other galaxies within those clusters \cite[][]{2018ApJ...862...95K, 2022ApJS..261...28Y, 2024arXiv240401560C}. Notably, the ICL behaves as a collisionless component, akin to dark matter, responding primarily to the cluster's global gravitational potential. This makes it a particularly promising tracer of the dark matter distribution in galaxy clusters \citep{Yoo2024, Yoo2025}.

Given the instrument's characteristics and resolution constraints, the K-DRIFT LSB imaging survey will prioritize the study of UDGs in different environments of the nearby universe as its primary mission. Their large size and low stellar density challenge conventional galaxy formation theories, raising questions about whether they are failed normal galaxies or extended dwarf galaxies shaped by environmental factors \cite[][]{2023MNRAS.522.1033B}. Conducting deep imaging of these galaxies offers a distinctive perspective on the formation and evolution of LSB systems. Discoveries in various environments, from clusters to voids, will allow for the examination of their structural properties, internal kinematics, and dark matter content, providing valuable insights into galaxy formation mechanisms. Specifically, studying UDGs within clusters will enable us to explore the impact of environmental processes, such as ram pressure stripping and tidal heating, on their development. Furthermore, we anticipate that an in-depth investigation of UDG properties in cosmic void regions will significantly advance our understanding of their origins.

\subsection{Advancement of the $\Lambda$CDM Model}
The $\Lambda$CDM model's well-known small-scale ($<$1 Mpc) issues, such as the Missing Satellites, Core/Cusp, and Too-Big-To-Fail problems, arise from discrepancies between cold dark matter (CDM) simulations and observed dwarf galaxies \cite[see][]{2017ARA&A..55..343B}. These challenges are being addressed both theoretically, through advanced hydrodynamic simulations that examine the influence of baryons \cite[][]{2022NatAs...6..897S}, and observationally, by expanding galaxy samples to improve statistical reliability. Nonetheless, limitations in photometry continue to impede these efforts, highlighting the need for more comprehensive samples. Consequently, a key objective for the next decade is to conduct deep imaging surveys in the nearby universe to identify and characterize faint, low-mass halos. These surveys will either reinforce the $\Lambda$CDM framework or necessitate a significant revision of our understanding of dark matter. Furthermore, upcoming high-resolution surveys (e.g., LSST) of distant dwarf galaxies hold great promise for future solutions to these small-scale issues.
 
\subsection{Transients, Variable Sources, and Stellar Substructures in the Milky Way} 
Given its wide FoV, K-DRIFT also offers the potential to study variable sources and stellar substructures in and around the Milky Way. Stellar variability plays a crucial role in understanding the physical properties of stars. Variable stars such as Cepheids and supernovae have been essential for measuring cosmic distances~\citep{Hubble1929} and for probing the expansion of the universe~\citep{Perlmutter1999,Riess1998}. Furthermore, populations such as RR Lyrae, Cepheids, and red clump stars provide key insights into the structure and formation history of galaxies. The $Gaia$ mission recently classified 12.4 million variable stars~\citep{Rimold2023}, but its sparse temporal sampling has led to some misclassifications. K-DRIFT can complement Gaia by observing a variety of variable and transient objects including AM CVn systems, Cepheids, RR Lyrae stars, and supernovae with higher cadence and broader coverage.
For instance, we plan to monitor 1,478 cataclysmic variables with a cadence of a few minutes. For RR Lyrae and Cepheid variables, we have identified 11,309 candidates with low classification scores as potential targets for monitoring with a cadence of several hours.

According to the $\Lambda$CDM model, the Milky Way was assembled through the hierarchical merging and accretion of low-mass fragments like dwarf galaxies
~\citep{Baugh1996,Diemand2007}. Indeed, nearly 100 stellar substructures and stellar streams have already been found in the Milky Way~\citep{Ibata1994,Malhan2022}. These substructures contain the fossil records of the formation of the Milky Way and provide important constraints on the distribution of dark matter. Many more undetected substructures of the Milky Way likely exist around intact satellite galaxies~\citep{Shipp2023}. The K-DRIFT survey will enable the detection of these substructures and help investigate the assembly history of the Milky Way.

\subsection{Circumgalactic Medium and Environmental Effects}
Galactic cirrus is one of the primary obstacles to detecting extragalactic LSB sources. This necessitates investigating the effect of the cirrus cloud features. Furthermore, interstellar dust clouds themselves are key objects for understanding many interesting issues, including dust properties and the interstellar radiation field, both of which are critical to studies of chemical evolution and star formation in galaxies. The K-DRIFT’s broad FoV and accurate background subtraction methods will be instrumental in measuring the reddening of background sources. This capability is expected to enable the effective detection of extended dust in galactic halos, the circumgalactic medium, and intra-cluster space.

The H$\alpha$ emission line provides valuable information about star formation activity in and around galaxies, as well as the environmental effects acting on them. For instance, the halo H$\alpha$ emission from diffuse ionized gas has been found to correlate well with dust-scattered UV emission in 38 nearby edge-on late-type galaxies \citep[e.g.,][]{Jo2018}. In galaxy clusters, strong ram pressure can directly strip the interstellar medium (ISM) from galaxies. The stripped ISM is then dissociated and ionized by the hot intracluster medium (ICM), forming ram pressure stripping tails that emit H$\alpha$ photons~\citep[e.g.,][]{gavazzi01,cortese06,yagi10,fumagalli14,yagi17}. Furthermore, when a gas-rich galaxy experiences strong ram pressure, the stripped ISM is mixed with the ICM, producing numerous warm ionized clouds that may cool and collapse into cold clouds before tails are destroyed~\citep{lee22}. In such condition, stars can form in the tail clumps, which also emit H$\alpha$ photons. The H$\alpha$ survey which will be conducted by K-DRIFT will enable us to investigate the evolutionary stage of ram pressure stripped galaxies in cluster environments.

\section{K-DRIFT Prospects} \label{sec:prospects}
\subsection{Legacy Imaging}
\subsubsection{The Role of Legacy Imaging in LSB Research}
The faint nature of LSB structures requires extreme mass and spatial resolution in cosmological simulations to model tidal debris and extended halos without excessive numerical diffusion. To address this computational challenge, we developed the Galaxy Replacement Technique \cite[GRT;][]{2022ApJ...925..103C}, enabling efficient modeling tidal stripping and satellite-host interactions at a mass resolution of $M_{\text{star}} = 7.9 \times 10^4 \, M_\odot \,$ and a spatial resolution of $\sim$15 pc. Our simulations are optimized to probe galaxy interactions and the formation of LSB systems within dense environments (e.g., galaxy clusters).

A consistent LSB imaging survey reaching a surface brightness of $\sim$30 mag arcsec$^{-2}$ over $\sim$5000 deg$^{2}$ provides a empirical baseline required to evaluate and improve theoretical models.
The integration of our GRT simulation framework with the extensive deep, wide-field legacy datasets from K-DRIFT enables a more rigorous comparison between simulated results and observed surface–brightness distributions, supporting comprehensive tests of the evolutionary properties of galaxies and galaxy clusters. 
Recent studies \cite[][]{Chun24,Chun25} have further developed this approach by producing surface brightness maps matched to the K-DRIFT specifications, demonstrating direct applicability to the project’s scientific objectives.
Accordingly, the availability of K-DRIFT's legacy archive is crucial both for validating our simulation predictions and for advancing our understanding of LSB galaxy formation and evolution.

\subsubsection{Complementary Multi-Wavelength Legacy Datasets}
Beyond its primary data collection, the K-DRIFT project leverages a wealth of complementary legacy datasets across various wavelengths to expand the scope of LSB research. This multi-wavelength approach provides a comprehensive perspective on LSB systems, from their stellar content to their gas dynamics.

Extensive optical and infrared surveys, such as LSST, SPHEREx~\citep{2020SPIE11443E..0IC}, and the DESI Legacy Imaging Surveys~\citep{2019AJ....157..168D}, provide valuable contextual data. This information, when combined with K-DRIFT's deep observations, facilitates statistical studies of faint objects across diverse environments and helps characterize the distribution and properties of LSB galaxies relative to their surroundings.

K-DRIFT imaging data will provide a crucial complement to integral-field-unit (IFU) spectroscopic surveys, such as HECTOR~\citep{2015IAUS..309...21B}, enabling a detailed investigation into the dynamical and stellar population properties of LSB objects.
These combined datasets offer spatially resolved kinematic and stellar population information, which will allow for a comprehensive characterization of LSB galaxies and the reconstruction of their evolutionary histories. Furthermore, high-resolution spectroscopic follow-up with facilities like existing large telescopes and the forthcoming Giant Magellan Telescope is expected to provide deeper insight into the physical conditions within LSB systems.

Radio survey data are also essential for tracing the neutral hydrogen content in LSB galaxies. While early large-scale HI surveys like the HI Parkes All Sky Survey \cite[HIPASS;][]{2001MNRAS.322..486B} provide a valuable low-resolution reference for overall gas distribution, newer facilities are revolutionizing this field. Deeper observations from the Australian Square Kilometre Array Pathfinder \cite[ASKAP;][]{2021PASA...38....9H} and MeerKAT \cite[][]{2018ApJ...856..180C} offer significantly higher sensitivity and spatial resolution. These datasets, along with future observations from the Square Kilometre Array \cite[SKA;][]{2009IEEEP..97.1482D}, will enable us to investigate the gas content of LSB systems across various environments, which is fundamental to understanding their formation and evolution.

In summary, by integrating deep imaging from K-DRIFT with multi-wavelength legacy datasets and LSB-optimized numerical simulations, we will establish a robust framework for a novel observational perspective on the LSB universe. This coordinated initiative will be critical for uncovering the formation and evolutionary pathways of LSB systems and understanding their significance within the broader context of cosmic evolution.

\subsection{Beyond Ground-Based Observations: Space-Based K-DRIFT}
Ground-based LSB imaging is fundamentally limited by atmospheric conditions and fluctuations in sky brightness. A space-based mission would mitigate these constraints by operating in a stable, low-background environment, thus enabling the achievement of unprecedented photometric depths that cannot be achieved from the ground. Additionally, it would not only be capable of observing sky regions that are difficult to access from specific ground-based facilities but also of access wavelengths that are obstructed from ground-based observations.

The ability to observe ultraviolet (UV) light is essential for investigating recent star formation activities. In the context of the nearby universe, it is widely accepted that most LSB objects, such as UDGs, tidal features, and ICL, are predominantly composed of older stellar populations \citep[e.g.,][]{2017ApJ...834...16M, Gu20}. However, a small proportion of young stars, which are not discernible at optical wavelengths but can be detected in the UV \cite[e.g.,][]{Yi+05,Ko+13,Ko+16}, provides valuable insight into the formation processes of these LSB objects.

Leveraging the experience gained from developing and operating the ground-based K-DRIFT, we propose a next-generation space telescope to undertake an all-sky survey of ultra–low-surface-brightness (uLSB; fainter than $\sim31$ mag arcsec$^{-2}$) imaging from the near-UV through the optical (200--700 nm). The mission is designed to reach a surface-brightness sensitivity at least one magnitude deeper than ground-based facilities in the optical (see Figure \ref{fig:f1}). Unlike the current ground-based K-DRIFT G1, which is limited to the southern hemisphere, the space telescope will enable uLSB observations across the entire sky with near-UV access. By combining full-sky coverage with superior sensitivity, this dedicated mission will substantially expand the census of diffuse structures and improve statistical robustness beyond what is possible with K-DRIFT G1.

A space-based K-DRIFT mission would serve as a crucial complement to the ongoing and upcoming ground-based surveys, enabling a powerful, synergistic approach to LSB research. This collaboration, taking advantage of a wider wavelength range and careful calibration, is expected to reveal the properties of uLSB systems that are currently beyond detection limits.

\section{Summary} \label{sec:remark}

The K-DRIFT project aims to tackle the observational challenges associated with LSB astronomical phenomena. These faint structures, including stellar streams, diffuse galaxy halos, and extended filaments, are crucial for understanding galaxy formation and evolution, especially within the $\Lambda$CDM cosmological model. However, their detection is severely hindered by their inherent faintness, stray light interference, and variations in the sky background, rendering traditional imaging methods insufficient for thorough investigations.

Studying LSB objects is crucial because they provide direct observational evidence of galaxy assembly history, interactions, and dark matter distribution. Although there have been numerous efforts in LSB imaging, the Universe below $\sim$29 mag arcsec$^{-2}$ remains largely unexplored. This observational gap represents a significant limitation that could bias our understanding of galaxy formation processes and cosmic evolution.

To address this issue, the K-DRIFT project has been developed to provide specialized observational solutions specifically for LSB imaging. The first generation telescope, K-DRIFT G1, has been successfully completed, incorporating advanced technologies to enhance the detection of LSB structures. Through the use of novel optical designs (off-axis freeform three-mirror), specialized imaging techniques (rolling dithering and dark-sky flat-fielding), and carefully planned observation strategies, K-DRIFT aims to achieve exceptional sensitivity and coverage, allowing the exploration of areas previously undetectable in wide-area surveys.

Looking forward, K-DRIFT's deep and wide survey in the southern hemisphere will enhance the worth of existing survey data and offer an early glimpse into the $\sim$ 30 mag arcsec$^{-2}$ universe, preceding the full 10-yr LSST dataset. This effort will foster significant synergies with LSST, leveraging shared observational strengths to vastly expand our comprehensive view of the LSB universe. 
Furthermore, the project is advancing toward the development of a dedicated space telescope optimized for ultra-low-surface-brightness, and designed to overcome the limitations of ground-based observations and lead to groundbreaking discoveries.
Ultimately, K-DRIFT aims not only to overcome observational difficulties but also to advance cosmological theories by providing crucial observational constraints that inform our understanding of galaxy evolution and the large-scale structure of the universe.


\acknowledgments
We are grateful to the anonymous reviewer for the helpful feedback, which enabled us to enhance the manuscript.
This research was supported by the Korea Astronomy and Space Science Institute under the R\&D program (Project No. 2025-1-831-00) supervised by the Ministry of Science and ICT.
J.Y. was supported by a KIAS Individual Grant (QP089902) via the Quantum Universe Center at Korea Institute for Advanced Study.


\appendix
\section{Calculation of Surface Brightness Limit} \label{sec:app1}
To fairly compare the photometric performance across various survey datasets---particularly for extended LSB structures---we focus on comparing their surface brightness limits. However, since surveys often report only the point-source detection limits, it is necessary to convert theses into a corresponding surface brightness limits. Below, we describe the step-by-step procedure for converting a 5$\sigma$ point-source detection limit to a 3$\sigma$ surface brightness limit. 

The point-source detection limit is calculated from the standard deviation of the ``total'' background signal integrated over the aperture, whereas the surface brightness limit is determined from the standard deviation of the ``mean'' background signal per pixel, averaged over a specified angular scale (e.g., $10^{\prime\prime}\times10^{\prime\prime}$) and expressed in units of arcsec$^{-2}$. The 5$\sigma$ detection limit for a point source can then be defined as:
\begin{equation}
m_\mathrm{limit} = \mathrm{ZP} - 2.5\ \log\left(5\sigma\sqrt{N}\right),
\end{equation}
where ZP is the photometric zero point, $\sigma$ is the standard deviation of the background per pixel (i.e., the root-mean-square value of the pixel-to-pixel variation in the background signal), and $N$ is the number of pixels within the aperture used for photometry. If a circular aperture with radius $r$ (in arcseconds) is used, and the pixel scale is $\mathrm{pix}$ (in arcseconds per pixel), then the number of pixels is computed as:
\begin{equation} \label{eq:2}
N = \pi \left(\frac{r}{\mathrm{pix}}\right)^2. 
\end{equation}
Here, the radius $r$ typically corresponds to the PSF FWHM or a chosen aperture radius, depending on the photometry method. Note that the detection limit defined here represents a nominal 5$\sigma$ sensitivity based solely on background noise. Therefore, it may differ slightly from the detection limit when source detection completeness is taken into account. 

The 3$\sigma$ level of the background signal per pixel, averaged over an angular scale of $10^{\prime\prime}\times10^{\prime\prime}$, is given by $3\sigma\times(\mathrm{pix}/10)$, according to the uncertainty propagation rule. This value is then multiplied by $1/\mathrm{pix}^2$ to convert it into units of arcsec$^{-2}$. Finally, the resulting surface brightness limit, measured within $10^{\prime\prime}\times10^{\prime\prime}$ boxes and expressed in units of mag arcsec$^{-2}$, is given by \cite[see also][]{2020A&A...644A..42R}:
\begin{equation}
\mu_\mathrm{limit} = \mathrm{ZP} - 2.5\ \log\left(\frac{3\sigma}{\mathrm{pix}\times10}\right),
\end{equation}
where, again, $\sigma$ is the standard deviation of the background per pixel, and $\mathrm{pix}$ is the pixel scale in arcseconds. This equation can be rewritten as:
\begin{equation}
\mu_\mathrm{limit} = \mathrm{ZP} - 2.5\ \log\left(5\sigma\sqrt{N}\times\frac{3}{\mathrm{pix}\times50\sqrt{N}}\right),
\end{equation}
and then simplified to,
\begin{equation}
\mu_\mathrm{limit} = m_\mathrm{limit} - 2.5\ \log\left(\frac{3}{\mathrm{pix}\times50\sqrt{N}}\right).
\end{equation}
Substituting the previously defined $N$ from equation (\ref{eq:2}), we finally obtain:
\begin{equation}
\mu_\mathrm{limit} = m_\mathrm{limit} - 2.5\ \log\left(\frac{3}{r\times 50\sqrt{\pi}}\right).
\end{equation}

Thus, the conversion from the point-source detection limit to the surface brightness limit can be directly derived based on the radius of the aperture used for photometry. This enables a straightforward comparison of surface brightness limits across different surveys using readily available parameters.



\begin{thebibliography}{}
\expandafter\ifx\csname natexlab\endcsname\relax\def\natexlab#1{#1}\fi
\providecommand{\url}[1]{\href{#1}{#1}}
\providecommand{\dodoi}[1]{doi:~\href{http://doi.org/#1}{\nolinkurl{#1}}}
\providecommand{\doeprint}[1]{\href{http://ascl.net/#1}{\nolinkurl{http://ascl.net/#1}}}
\providecommand{\doarXiv}[1]{\href{https://arxiv.org/abs/#1}{\nolinkurl{https://arxiv.org/abs/#1}}}
\providecommand{\dodoilink}[2]{\href{http://doi.org/#1}{#2}}
\providecommand{\doadslink}[2]{\href{#1}{#2}}

\bibitem[{{Abraham} \& {van Dokkum}(2014)}]{2014PASP..126...55A}
{Abraham}, R.~G., \& {van Dokkum}, P.~G. 2014, \pasp, 126, 55

\bibitem[{{Aihara} {et~al.}(2018){Aihara}, {Arimoto}, {Armstrong}, {Arnouts}, {Bahcall}, {Bickerton}, {Bosch}, {Bundy}, {Capak}, {Chan}, {Chiba}, {Coupon}, {Egami}, {Enoki}, {Finet}, {Fujimori}, {Fujimoto}, {Furusawa}, {Furusawa}, {Goto}, {Goulding}, {Greco}, {Greene}, {Gunn}, {Hamana}, {Harikane}, {Hashimoto}, {Hattori}, {Hayashi}, {Hayashi}, {He{\l}miniak}, {Higuchi}, {Hikage}, {Ho}, {Hsieh}, {Huang}, {Huang}, {Ikeda}, {Imanishi}, {Inoue}, {Iwasawa}, {Iwata}, {Jaelani}, {Jian}, {Kamata}, {Karoji}, {Kashikawa}, {Katayama}, {Kawanomoto}, {Kayo}, {Koda}, {Koike}, {Kojima}, {Komiyama}, {Konno}, {Koshida}, {Koyama}, {Kusakabe}, {Leauthaud}, {Lee}, {Lin}, {Lin}, {Lupton}, {Mandelbaum}, {Matsuoka}, {Medezinski}, {Mineo}, {Miyama}, {Miyatake}, {Miyazaki}, {Momose}, {More}, {More}, {Moritani}, {Moriya}, {Morokuma}, {Mukae}, {Murata}, {Murayama}, {Nagao}, {Nakata}, {Niida}, {Niikura}, {Nishizawa}, {Obuchi}, {Oguri}, {Oishi}, {Okabe}, {Okamoto}, {Okura}, {Ono}, {Onodera}, {Onoue}, {Osato}, {Ouchi}, {Price}, {Pyo},
  {Sako}, {Sawicki}, {Shibuya}, {Shimasaku}, {Shimono}, {Shirasaki}, {Silverman}, {Simet}, {Speagle}, {Spergel}, {Strauss}, {Sugahara}, {Sugiyama}, {Suto}, {Suyu}, {Suzuki}, {Tait}, {Takada}, {Takata}, {Tamura}, {Tanaka}, {Tanaka}, {Tanaka}, {Tanaka}, {Terai}, {Terashima}, {Toba}, {Tominaga}, {Toshikawa}, {Turner}, {Uchida}, {Uchiyama}, {Umetsu}, {Uraguchi}, {Urata}, {Usuda}, {Utsumi}, {Wang}, {Wang}, {Wong}, {Yabe}, {Yamada}, {Yamanoi}, {Yasuda}, {Yeh}, {Yonehara}, \& {Yuma}}]{2018PASJ...70S...4A}
{Aihara}, H., {Arimoto}, N., {Armstrong}, R., {et~al.} 2018, \pasj, 70, S4

\bibitem[{{Aihara} {et~al.}(2022){Aihara}, {AlSayyad}, {Ando}, {Armstrong}, {Bosch}, {Egami}, {Furusawa}, {Furusawa}, {Harasawa}, {Harikane}, {Hsieh}, {Ikeda}, {Ito}, {Iwata}, {Kodama}, {Koike}, {Kokubo}, {Komiyama}, {Li}, {Liang}, {Lin}, {Lupton}, {Lust}, {MacArthur}, {Mawatari}, {Mineo}, {Miyatake}, {Miyazaki}, {More}, {Morishima}, {Murayama}, {Nakajima}, {Nakata}, {Nishizawa}, {Oguri}, {Okabe}, {Okura}, {Ono}, {Osato}, {Ouchi}, {Pan}, {Plazas Malag{\'o}n}, {Price}, {Reed}, {Rykoff}, {Shibuya}, {Simunovic}, {Strauss}, {Sugimori}, {Suto}, {Suzuki}, {Takada}, {Takagi}, {Takata}, {Takita}, {Tanaka}, {Tang}, {Taranu}, {Terai}, {Toba}, {Turner}, {Uchiyama}, {Vijarnwannaluk}, {Waters}, {Yamada}, {Yamamoto}, \& {Yamashita}}]{2022PASJ...74..247A}
{Aihara}, H., {AlSayyad}, Y., {Ando}, M., {et~al.} 2022, \pasj, 74, 247

\bibitem[{{Barnes} {et~al.}(2001){Barnes}, {Staveley-Smith}, {de Blok}, {Oosterloo}, {Stewart}, {Wright}, {Banks}, {Bhathal}, {Boyce}, {Calabretta}, {Disney}, {Drinkwater}, {Ekers}, {Freeman}, {Gibson}, {Green}, {Haynes}, {te Lintel Hekkert}, {Henning}, {Jerjen}, {Juraszek}, {Kesteven}, {Kilborn}, {Knezek}, {Koribalski}, {Kraan-Korteweg}, {Malin}, {Marquarding}, {Minchin}, {Mould}, {Price}, {Putman}, {Ryder}, {Sadler}, {Schr{\"o}der}, {Stootman}, {Webster}, {Wilson}, \& {Ye}}]{2001MNRAS.322..486B}
{Barnes}, D.~G., {Staveley-Smith}, L., {de Blok}, W.~J.~G., {et~al.} 2001, \mnras, 322, 486

\bibitem[{{Baugh} {et~al.}(1996){Baugh}, {Cole}, \& {Frenk}}]{Baugh1996}
{Baugh}, C.~M., {Cole}, S., \& {Frenk}, C.~S. 1996, \mnras, 283, 1361

\bibitem[{{Beckwith} {et~al.}(2006){Beckwith}, {Stiavelli}, {Koekemoer}, {Caldwell}, {Ferguson}, {Hook}, {Lucas}, {Bergeron}, {Corbin}, {Jogee}, {Panagia}, {Robberto}, {Royle}, {Somerville}, \& {Sosey}}]{2006AJ....132.1729B}
{Beckwith}, S. V.~W., {Stiavelli}, M., {Koekemoer}, A.~M., {et~al.} 2006, \aj, 132, 1729

\bibitem[{{Belokurov} {et~al.}(2007){Belokurov}, {Evans}, {Irwin}, {Lynden-Bell}, {Yanny}, {Vidrih}, {Gilmore}, {Seabroke}, {Zucker}, {Wilkinson}, {Hewett}, {Bramich}, {Fellhauer}, {Newberg}, {Wyse}, {Beers}, {Bell}, {Barentine}, {Brinkmann}, {Cole}, {Pan}, \& {York}}]{2007ApJ...658..337B}
{Belokurov}, V., {Evans}, N.~W., {Irwin}, M.~J., {et~al.} 2007, \apj, 658, 337

\bibitem[{{Benavides} {et~al.}(2023){Benavides}, {Sales}, {Abadi}, {Marinacci}, {Vogelsberger}, \& {Hernquist}}]{2023MNRAS.522.1033B}
{Benavides}, J.~A., {Sales}, L.~V., {Abadi}, M.~G., {et~al.} 2023, \mnras, 522, 1033

\bibitem[{{Bland-Hawthorn}(2015)}]{2015IAUS..309...21B}
{Bland-Hawthorn}, J. 2015, in IAU Symposium, Vol. 309, Galaxies in 3D across the Universe, ed. B.~L. {Ziegler}, F.~{Combes}, H.~{Dannerbauer}, \& M.~{Verdugo}, 21--28, \dodoi{10.1017/S1743921314009247}

\bibitem[{{Borlaff} \& {Marcum}(2022)}]{2022AAS...24030406B}
{Borlaff}, A., \& {Marcum}, P. 2022, in American Astronomical Society Meeting Abstracts, Vol. 240, American Astronomical Society Meeting \#240, 304.06

\bibitem[{{Borlaff} {et~al.}(2019){Borlaff}, {Trujillo}, {Rom{\'a}n}, {Beckman}, {Eliche-Moral}, {Infante-S{\'a}inz}, {Lumbreras-Calle}, {de Almagro}, {G{\'o}mez-Guijarro}, {Cebri{\'a}n}, {Dorta}, {Cardiel}, {Akhlaghi}, \& {Mart{\'\i}nez-Lombilla}}]{2019A&A...621A.133B}
{Borlaff}, A., {Trujillo}, I., {Rom{\'a}n}, J., {et~al.} 2019, \aap, 621, A133

\bibitem[{{Bullock} \& {Boylan-Kolchin}(2017)}]{2017ARA&A..55..343B}
{Bullock}, J.~S., \& {Boylan-Kolchin}, M. 2017, \araa, 55, 343

\bibitem[{{Byun} {et~al.}(2025){Byun}, {Seon}, \& {Ko}}]{2025PASP..137e4502B}
{Byun}, W., {Seon}, K.-I., \& {Ko}, J. 2025, \pasp, 137, 054502

\bibitem[{{Byun} {et~al.}(2022){Byun}, {Ko}, {Kim}, {Seon}, {Chang}, {Kim}, {Choi}, {Chun}, {Jeon}, {Kim}, {Lee}, {Lee}, {Park}, {Sung}, {Yoo}, {Lee}, \& {Lee}}]{2022PASP..134h4101B}
{Byun}, W., {Ko}, J., {Kim}, Y., {et~al.} 2022, \pasp, 134, 084101

\bibitem[{{Camilo} {et~al.}(2018){Camilo}, {Scholz}, {Serylak}, {Buchner}, {Merryfield}, {Kaspi}, {Archibald}, {Bailes}, {Jameson}, {van Straten}, {Sarkissian}, {Reynolds}, {Johnston}, {Hobbs}, {Abbott}, {Adam}, {Adams}, {Alberts}, {Andreas}, {Asad}, {Baker}, {Baloyi}, {Bauermeister}, {Baxana}, {Bennett}, {Bernardi}, {Booisen}, {Booth}, {Botha}, {Boyana}, {Brederode}, {Burger}, {Cheetham}, {Conradie}, {Conradie}, {Davidson}, {De Bruin}, {de Swardt}, {de Villiers}, {de Villiers}, {de Villiers}, {de Villiers}, {De Waal}, {Dikgale}, {du Toit}, {du Toit}, {Esterhuyse}, {Fanaroff}, {Fataar}, {Foley}, {Foster}, {Fourie}, {Gamatham}, {Gatsi}, {Geschke}, {Goedhart}, {Grobler}, {Gumede}, {Hlakola}, {Hokwana}, {Hoorn}, {Horn}, {Horrell}, {Hugo}, {Isaacson}, {Jacobs}, {Jansen van Rensburg}, {Jonas}, {Jordaan}, {Joubert}, {Joubert}, {J{\'o}zsa}, {Julie}, {Julius}, {Kapp}, {Karastergiou}, {Karels}, {Kariseb}, {Karuppusamy}, {Kasper}, {Knox-Davies}, {Koch}, {Kotz{\'e}}, {Krebs}, {Kriek}, {Kriel}, {Kusel}, {Lamoor},
  {Lehmensiek}, {Liebenberg}, {Liebenberg}, {Lord}, {Lunsky}, {Mabombo}, {Macdonald}, {Macfarlane}, {Madisa}, {Mafhungo}, {Magnus}, {Magozore}, {Mahgoub}, {Main}, {Makhathini}, {Malan}, {Malgas}, {Manley}, {Manzini}, {Marais}, {Marais}, {Marais}, {Maree}, {Martens}, {Matshawule}, {Matthysen}, {Mauch}, {McNally}, {Merry}, {Millenaar}, {Mjikelo}, {Mkhabela}, {Mnyandu}, {Moeng}, {Mokone}, {Monama}, {Montshiwa}, {Moss}, {Mphego}, {New}, {Ngcebetsha}, {Ngoasheng}, {Niehaus}, {Ntuli}, {Nzama}, {Obies}, {Obrocka}, {Ockards}, {Olyn}, {Oozeer}, {Otto}, {Padayachee}, {Passmoor}, {Patel}, {Paula}, {Peens-Hough}, {Pholoholo}, {Prozesky}, {Rakoma}, {Ramaila}, {Rammala}, {Ramudzuli}, {Rasivhaga}, {Ratcliffe}, {Reader}, {Renil}, {Richter}, {Robyntjies}, {Rosekrans}, {Rust}, {Salie}, {Sambu}, {Schollar}, {Schwardt}, {Seranyane}, {Sethosa}, {Sharpe}, {Siebrits}, {Sirothia}, {Slabber}, {Smirnov}, {Smith}, {Sofeya}, {Songqumase}, {Spann}, {Stappers}, {Steyn}, {Steyn}, {Strong}, {Struthers}, {Stuart}, {Sunnylall}, {Swart},
  {Taljaard}, {Tasse}, {Taylor}, {Theron}, {Thondikulam}, {Thorat}, {Tiplady}, {Toruvanda}, {van Aardt}, {van Balla}, {van den Heever}, {van der Byl}, {van der Merwe}, {van der Merwe}, {van Niekerk}, {van Rooyen}, {van Staden}, {van Tonder}, \& {van Wyk}}]{2018ApJ...856..180C}
{Camilo}, F., {Scholz}, P., {Serylak}, M., {et~al.} 2018, \apj, 856, 180

\bibitem[{{Carollo} {et~al.}(2007){Carollo}, {Beers}, {Lee}, {Chiba}, {Norris}, {Wilhelm}, {Sivarani}, {Marsteller}, {Munn}, {Bailer-Jones}, {Fiorentin}, \& {York}}]{2007Natur.450.1020C}
{Carollo}, D., {Beers}, T.~C., {Lee}, Y.~S., {et~al.} 2007, \nat, 450, 1020

\bibitem[{Chang(2013)}]{10.1117/12.2023433}
Chang, S. 2013, in UV/Optical/IR Space Telescopes and Instruments: Innovative Technologies and Concepts VI, ed. H.~A. MacEwen \& J.~B. Breckinridge, Vol. 8860, International Society for Optics and Photonics (SPIE), 219 -- 228, \dodoi{10.1117/12.2023433}

\bibitem[{{Cheriyan} {et~al.}(2022){Cheriyan}, {Segovia de la Torre}, {Villegas Calvo}, {Kurvits}, {Nottingham}, \& {McClure}}]{2022SPIE12191E..0IC}
{Cheriyan}, S., {Segovia de la Torre}, J.~A., {Villegas Calvo}, J.~A., {et~al.} 2022, in Society of Photo-Optical Instrumentation Engineers (SPIE) Conference Series, Vol. 12191, X-Ray, Optical, and Infrared Detectors for Astronomy X, ed. A.~D. {Holland} \& J.~{Beletic}, 121910I, \dodoi{10.1117/12.2634291}

\bibitem[{{Chun} {et~al.}(2025){Chun}, {Shin}, {Ko}, {Smith}, {Park}, \& {Nam}}]{Chun25}
{Chun}, K., {Shin}, J., {Ko}, J., {et~al.} 2025, arXiv e-prints, arXiv:2509.22802

\bibitem[{{Chun} {et~al.}(2024){Chun}, {Shin}, {Ko}, {Smith}, \& {Yoo}}]{Chun24}
{Chun}, K., {Shin}, J., {Ko}, J., {Smith}, R., \& {Yoo}, J. 2024, \apj, 969, 142

\bibitem[{{Chun} {et~al.}(2022){Chun}, {Shin}, {Smith}, {Ko}, \& {Yoo}}]{2022ApJ...925..103C}
{Chun}, K., {Shin}, J., {Smith}, R., {Ko}, J., \& {Yoo}, J. 2022, \apj, 925, 103

\bibitem[{{Contini} {et~al.}(2024){Contini}, {Yi}, \& {Jeon}}]{2024arXiv240401560C}
{Contini}, E., {Yi}, S.~K., \& {Jeon}, S. 2024, arXiv e-prints, arXiv:2404.01560

\bibitem[{{Cooper} {et~al.}(2013){Cooper}, {D'Souza}, {Kauffmann}, {Wang}, {Boylan-Kolchin}, {Guo}, {Frenk}, \& {White}}]{2013MNRAS.434.3348C}
{Cooper}, A.~P., {D'Souza}, R., {Kauffmann}, G., {et~al.} 2013, \mnras, 434, 3348

\bibitem[{{Cortese} {et~al.}(2006){Cortese}, {Gavazzi}, {Boselli}, {Franzetti}, {Kennicutt}, {O'Neil}, \& {Sakai}}]{cortese06}
{Cortese}, L., {Gavazzi}, G., {Boselli}, A., {et~al.} 2006, \aap, 453, 847

\bibitem[{{Crill} {et~al.}(2020){Crill}, {Werner}, {Akeson}, {Ashby}, {Bleem}, {Bock}, {Bryan}, {Burnham}, {Byunh}, {Chang}, {Chiang}, {Cook}, {Cooray}, {Davis}, {Dor{\'e}}, {Dowell}, {Dubois-Felsmann}, {Eifler}, {Faisst}, {Habib}, {Heinrich}, {Heitmann}, {Heaton}, {Hirata}, {Hristov}, {Hui}, {Jeong}, {Kang}, {Kecman}, {Kirkpatrick}, {Korngut}, {Krause}, {Lee}, {Lisse}, {Masters}, {Mauskopf}, {Melnick}, {Miyasaka}, {Nayyeri}, {Nguyen}, {{\"O}berg}, {Padin}, {Paladini}, {Pourrahmani}, {Pyo}, {Smith}, {Song}, {Symons}, {Teplitz}, {Tolls}, {Unwin}, {Windhorst}, {Yang}, \& {Zemcov}}]{2020SPIE11443E..0IC}
{Crill}, B.~P., {Werner}, M., {Akeson}, R., {et~al.} 2020, in Society of Photo-Optical Instrumentation Engineers (SPIE) Conference Series, Vol. 11443, Space Telescopes and Instrumentation 2020: Optical, Infrared, and Millimeter Wave, ed. M.~{Lystrup} \& M.~D. {Perrin}, 114430I, \dodoi{10.1117/12.2567224}

\bibitem[{{Danieli} {et~al.}(2020){Danieli}, {Lokhorst}, {Zhang}, {Merritt}, {van Dokkum}, {Abraham}, {Conroy}, {Gilhuly}, {Greco}, {Janssens}, {Li}, {Liu}, {Miller}, \& {Mowla}}]{2020ApJ...894..119D}
{Danieli}, S., {Lokhorst}, D., {Zhang}, J., {et~al.} 2020, \apj, 894, 119

\bibitem[{{Dewdney} {et~al.}(2009){Dewdney}, {Hall}, {Schilizzi}, \& {Lazio}}]{2009IEEEP..97.1482D}
{Dewdney}, P.~E., {Hall}, P.~J., {Schilizzi}, R.~T., \& {Lazio}, T.~J.~L.~W. 2009, IEEE Proceedings, 97, 1482

\bibitem[{{Dey} {et~al.}(2019){Dey}, {Schlegel}, {Lang}, {Blum}, {Burleigh}, {Fan}, {Findlay}, {Finkbeiner}, {Herrera}, {Juneau}, {Landriau}, {Levi}, {McGreer}, {Meisner}, {Myers}, {Moustakas}, {Nugent}, {Patej}, {Schlafly}, {Walker}, {Valdes}, {Weaver}, {Y{\`e}che}, {Zou}, {Zhou}, {Abareshi}, {Abbott}, {Abolfathi}, {Aguilera}, {Alam}, {Allen}, {Alvarez}, {Annis}, {Ansarinejad}, {Aubert}, {Beechert}, {Bell}, {BenZvi}, {Beutler}, {Bielby}, {Bolton}, {Brice{\~n}o}, {Buckley-Geer}, {Butler}, {Calamida}, {Carlberg}, {Carter}, {Casas}, {Castander}, {Choi}, {Comparat}, {Cukanovaite}, {Delubac}, {DeVries}, {Dey}, {Dhungana}, {Dickinson}, {Ding}, {Donaldson}, {Duan}, {Duckworth}, {Eftekharzadeh}, {Eisenstein}, {Etourneau}, {Fagrelius}, {Farihi}, {Fitzpatrick}, {Font-Ribera}, {Fulmer}, {G{\"a}nsicke}, {Gaztanaga}, {George}, {Gerdes}, {Gontcho}, {Gorgoni}, {Green}, {Guy}, {Harmer}, {Hernandez}, {Honscheid}, {Huang}, {James}, {Jannuzi}, {Jiang}, {Joyce}, {Karcher}, {Karkar}, {Kehoe}, {Kneib}, {Kueter-Young}, {Lan},
  {Lauer}, {Le Guillou}, {Le Van Suu}, {Lee}, {Lesser}, {Perreault Levasseur}, {Li}, {Mann}, {Marshall}, {Mart{\'\i}nez-V{\'a}zquez}, {Martini}, {du Mas des Bourboux}, {McManus}, {Meier}, {M{\'e}nard}, {Metcalfe}, {Mu{\~n}oz-Guti{\'e}rrez}, {Najita}, {Napier}, {Narayan}, {Newman}, {Nie}, {Nord}, {Norman}, {Olsen}, {Paat}, {Palanque-Delabrouille}, {Peng}, {Poppett}, {Poremba}, {Prakash}, {Rabinowitz}, {Raichoor}, {Rezaie}, {Robertson}, {Roe}, {Ross}, {Ross}, {Rudnick}, {Safonova}, {Saha}, {S{\'a}nchez}, {Savary}, {Schweiker}, {Scott}, {Seo}, {Shan}, {Silva}, {Slepian}, {Soto}, {Sprayberry}, {Staten}, {Stillman}, {Stupak}, {Summers}, {Sien Tie}, {Tirado}, {Vargas-Maga{\~n}a}, {Vivas}, {Wechsler}, {Williams}, {Yang}, {Yang}, {Yapici}, {Zaritsky}, {Zenteno}, {Zhang}, {Zhang}, {Zhou}, \& {Zhou}}]{2019AJ....157..168D}
{Dey}, A., {Schlegel}, D.~J., {Lang}, D., {et~al.} 2019, \aj, 157, 168

\bibitem[{{Diemand} {et~al.}(2007){Diemand}, {Kuhlen}, \& {Madau}}]{Diemand2007}
{Diemand}, J., {Kuhlen}, M., \& {Madau}, P. 2007, \apj, 667, 859

\bibitem[{{Duc} {et~al.}(2015){Duc}, {Cuillandre}, {Karabal}, {Cappellari}, {Alatalo}, {Blitz}, {Bournaud}, {Bureau}, {Crocker}, {Davies}, {Davis}, {de Zeeuw}, {Emsellem}, {Khochfar}, {Krajnovi{\'c}}, {Kuntschner}, {McDermid}, {Michel-Dansac}, {Morganti}, {Naab}, {Oosterloo}, {Paudel}, {Sarzi}, {Scott}, {Serra}, {Weijmans}, \& {Young}}]{2015MNRAS.446..120D}
{Duc}, P.-A., {Cuillandre}, J.-C., {Karabal}, E., {et~al.} 2015, \mnras, 446, 120

\bibitem[{{Eigenthaler} {et~al.}(2018){Eigenthaler}, {Puzia}, {Taylor}, {Ordenes-Brice{\~n}o}, {Mu{\~n}oz}, {Ribbeck}, {Alamo-Mart{\'\i}nez}, {Zhang}, {{\'A}ngel}, {Capaccioli}, {C{\^o}t{\'e}}, {Ferrarese}, {Galaz}, {Grebel}, {Hempel}, {Hilker}, {Lan{\c{c}}on}, {Mieske}, {Miller}, {Paolillo}, {Powalka}, {Richtler}, {Roediger}, {Rong}, {S{\'a}nchez-Janssen}, \& {Spengler}}]{2018ApJ...855..142E}
{Eigenthaler}, P., {Puzia}, T.~H., {Taylor}, M.~A., {et~al.} 2018, \apj, 855, 142

\bibitem[{{Euclid Collaboration} {et~al.}(2022){Euclid Collaboration}, {Borlaff}, {G{\'o}mez-Alvarez}, {Altieri}, {Marcum}, {Vavrek}, {Laureijs}, {Kohley}, {Buitrago}, {Cuillandre}, {Duc}, {Gaspar Venancio}, {Amara}, {Andreon}, {Auricchio}, {Azzollini}, {Baccigalupi}, {Balaguera-Antol{\'\i}nez}, {Baldi}, {Bardelli}, {Bender}, {Biviano}, {Bodendorf}, {Bonino}, {Bozzo}, {Branchini}, {Brescia}, {Brinchmann}, {Burigana}, {Cabanac}, {Camera}, {Candini}, {Capobianco}, {Cappi}, {Carbone}, {Carretero}, {Carvalho}, {Casas}, {Castander}, {Castellano}, {Castignani}, {Cavuoti}, {Cimatti}, {Cledassou}, {Colodro-Conde}, {Congedo}, {Conselice}, {Conversi}, {Copin}, {Corcione}, {Coupon}, {Courtois}, {Cropper}, {Da Silva}, {Degaudenzi}, {Di Ferdinando}, {Douspis}, {Dubath}, {Duncan}, {Dupac}, {Dusini}, {Ealet}, {Fabricius}, {Farina}, {Farrens}, {Ferreira}, {Ferriol}, {Finelli}, {Flose-Reimberg}, {Fosalba}, {Frailis}, {Franceschi}, {Fumana}, {Galeotta}, {Ganga}, {Garilli}, {Gillis}, {Giocoli}, {Gozaliasl}, {Graci{\'a}-Carpio},
  {Grazian}, {Grupp}, {Haugan}, {Holmes}, {Hormuth}, {Jahnke}, {Keihanen}, {Kermiche}, {Kiessling}, {Kilbinger}, {Kirkpatrick}, {Kitching}, {Knapen}, {Kubik}, {K{\"u}mmel}, {Kunz}, {Kurki-Suonio}, {Liebing}, {Ligori}, {Lilje}, {Lindholm}, {Lloro}, {Mainetti}, {Maino}, {Mansutti}, {Marggraf}, {Markovic}, {Martinelli}, {Martinet}, {Mart{\'\i}nez-Delgado}, {Marulli}, {Massey}, {Maturi}, {Maurogordato}, {Medinaceli}, {Mei}, {Meneghetti}, {Merlin}, {Metcalf}, {Meylan}, {Moresco}, {Morgante}, {Moscardini}, {Munari}, {Nakajima}, {Neissner}, {Niemi}, {Nightingale}, {Nucita}, {Padilla}, {Paltani}, {Pasian}, {Patrizii}, {Pedersen}, {Percival}, {Pettorino}, {Pires}, {Poncet}, {Popa}, {Potter}, {Pozzetti}, {Raison}, {Rebolo}, {Renzi}, {Rhodes}, {Riccio}, {Romelli}, {Roncarelli}, {Rosset}, {Rossetti}, {Saglia}, {S{\'a}nchez}, {Sapone}, {Sauvage}, {Schneider}, {Scottez}, {Secroun}, {Seidel}, {Serrano}, {Sirignano}, {Sirri}, {Skottfelt}, {Stanco}, {Starck}, {Sureau}, {Tallada-Cresp{\'\i}}, {Taylor}, {Tenti}, {Tereno},
  {Teyssier}, {Toledo-Moreo}, {Torradeflot}, {Tutusaus}, {Valentijn}, {Valenziano}, {Valiviita}, {Vassallo}, {Viel}, {Wang}, {Weller}, {Whittaker}, {Zacchei}, {Zamorani}, \& {Zucca}}]{2022A&A...657A..92E}
{Euclid Collaboration}, {Borlaff}, A.~S., {G{\'o}mez-Alvarez}, P., {et~al.} 2022, \aap, 657, A92

\bibitem[{{Ferguson} {et~al.}(2002){Ferguson}, {Irwin}, {Ibata}, {Lewis}, \& {Tanvir}}]{2002AJ....124.1452F}
{Ferguson}, A. M.~N., {Irwin}, M.~J., {Ibata}, R.~A., {Lewis}, G.~F., \& {Tanvir}, N.~R. 2002, \aj, 124, 1452

\bibitem[{{Ferrarese} {et~al.}(2012){Ferrarese}, {C{\^o}t{\'e}}, {Cuillandre}, {Gwyn}, {Peng}, {MacArthur}, {Duc}, {Boselli}, {Mei}, {Erben}, {McConnachie}, {Durrell}, {Mihos}, {Jord{\'a}n}, {Lan{\c{c}}on}, {Puzia}, {Emsellem}, {Balogh}, {Blakeslee}, {van Waerbeke}, {Gavazzi}, {Vollmer}, {Kavelaars}, {Woods}, {Ball}, {Boissier}, {Courteau}, {Ferriere}, {Gavazzi}, {Hildebrandt}, {Hudelot}, {Huertas-Company}, {Liu}, {McLaughlin}, {Mellier}, {Milkeraitis}, {Schade}, {Balkowski}, {Bournaud}, {Carlberg}, {Chapman}, {Hoekstra}, {Peng}, {Sawicki}, {Simard}, {Taylor}, {Tully}, {van Driel}, {Wilson}, {Burdullis}, {Mahoney}, \& {Manset}}]{2012ApJS..200....4F}
{Ferrarese}, L., {C{\^o}t{\'e}}, P., {Cuillandre}, J.-C., {et~al.} 2012, \apjs, 200, 4

\bibitem[{{Fliri} \& {Trujillo}(2016)}]{2016MNRAS.456.1359F}
{Fliri}, J., \& {Trujillo}, I. 2016, \mnras, 456, 1359

\bibitem[{{Fumagalli} {et~al.}(2014){Fumagalli}, {Fossati}, {Hau}, {Gavazzi}, {Bower}, {Sun}, \& {Boselli}}]{fumagalli14}
{Fumagalli}, M., {Fossati}, M., {Hau}, G. K.~T., {et~al.} 2014, \mnras, 445, 4335

\bibitem[{{Gavazzi} {et~al.}(2001){Gavazzi}, {Boselli}, {Mayer}, {Iglesias-Paramo}, {V{\'\i}lchez}, \& {Carrasco}}]{gavazzi01}
{Gavazzi}, G., {Boselli}, A., {Mayer}, L., {et~al.} 2001, \apjl, 563, L23

\bibitem[{{Greco} {et~al.}(2018){Greco}, {Greene}, {Strauss}, {Macarthur}, {Flowers}, {Goulding}, {Huang}, {Kim}, {Komiyama}, {Leauthaud}, {Leisman}, {Lupton}, {Sif{\'o}n}, \& {Wang}}]{2018ApJ...857..104G}
{Greco}, J.~P., {Greene}, J.~E., {Strauss}, M.~A., {et~al.} 2018, \apj, 857, 104

\bibitem[{{Gu} {et~al.}(2020){Gu}, {Conroy}, {Law}, {van Dokkum}, {Yan}, {Wake}, {Bundy}, {Villaume}, {Abraham}, {Merritt}, {Zhang}, {Bershady}, {Bizyaev}, {Drory}, {Pan}, {Thomas}, \& {Weijmans}}]{Gu20}
{Gu}, M., {Conroy}, C., {Law}, D., {et~al.} 2020, \apj, 894, 32

\bibitem[{{Hernquist} \& {Spergel}(1992)}]{1992ApJ...399L.117H}
{Hernquist}, L., \& {Spergel}, D.~N. 1992, \apjl, 399, L117

\bibitem[{{Hotan} {et~al.}(2021){Hotan}, {Bunton}, {Chippendale}, {Whiting}, {Tuthill}, {Moss}, {McConnell}, {Amy}, {Huynh}, {Allison}, {Anderson}, {Bannister}, {Bastholm}, {Beresford}, {Bock}, {Bolton}, {Chapman}, {Chow}, {Collier}, {Cooray}, {Cornwell}, {Diamond}, {Edwards}, {Feain}, {Franzen}, {George}, {Gupta}, {Hampson}, {Harvey-Smith}, {Hayman}, {Heywood}, {Jacka}, {Jackson}, {Jackson}, {Jeganathan}, {Johnston}, {Kesteven}, {Kleiner}, {Koribalski}, {Lee-Waddell}, {Lenc}, {Lensson}, {Mackay}, {Mahony}, {McClure-Griffiths}, {McConigley}, {Mirtschin}, {Ng}, {Norris}, {Pearce}, {Phillips}, {Pilawa}, {Raja}, {Reynolds}, {Roberts}, {Roxby}, {Sadler}, {Shields}, {Schinckel}, {Serra}, {Shaw}, {Sweetnam}, {Troup}, {Tzioumis}, {Voronkov}, \& {Westmeier}}]{2021PASA...38....9H}
{Hotan}, A.~W., {Bunton}, J.~D., {Chippendale}, A.~P., {et~al.} 2021, \pasa, 38, e009

\bibitem[{{Hubble}(1929)}]{Hubble1929}
{Hubble}, E.~P. 1929, \apj, 69, 103

\bibitem[{{Ibata} {et~al.}(2001){Ibata}, {Irwin}, {Lewis}, {Ferguson}, \& {Tanvir}}]{2001Natur.412...49I}
{Ibata}, R., {Irwin}, M., {Lewis}, G., {Ferguson}, A. M.~N., \& {Tanvir}, N. 2001, \nat, 412, 49

\bibitem[{{Ibata} {et~al.}(2007){Ibata}, {Martin}, {Irwin}, {Chapman}, {Ferguson}, {Lewis}, \& {McConnachie}}]{2007ApJ...671.1591I}
{Ibata}, R., {Martin}, N.~F., {Irwin}, M., {et~al.} 2007, \apj, 671, 1591

\bibitem[{{Ibata} {et~al.}(1994){Ibata}, {Gilmore}, \& {Irwin}}]{Ibata1994}
{Ibata}, R.~A., {Gilmore}, G., \& {Irwin}, M.~J. 1994, \nat, 370, 194

\bibitem[{{Illingworth} {et~al.}(2013){Illingworth}, {Magee}, {Oesch}, {Bouwens}, {Labb{\'e}}, {Stiavelli}, {van Dokkum}, {Franx}, {Trenti}, {Carollo}, \& {Gonzalez}}]{2013ApJS..209....6I}
{Illingworth}, G.~D., {Magee}, D., {Oesch}, P.~A., {et~al.} 2013, \apjs, 209, 6

\bibitem[{{Iodice} {et~al.}(2021){Iodice}, {Spavone}, {Raj}, {Capaccioli}, {Cantiello}, \& {VEGAS science team}}]{2021arXiv210204950I}
{Iodice}, E., {Spavone}, M., {Raj}, M.~A., {et~al.} 2021, arXiv e-prints, arXiv:2102.04950

\bibitem[{{Ivezi{\'c}} {et~al.}(2019){Ivezi{\'c}}, {Kahn}, {Tyson}, {Abel}, {Acosta}, {Allsman}, {Alonso}, {AlSayyad}, {Anderson}, {Andrew}, {Angel}, {Angeli}, {Ansari}, {Antilogus}, {Araujo}, {Armstrong}, {Arndt}, {Astier}, {Aubourg}, {Auza}, {Axelrod}, {Bard}, {Barr}, {Barrau}, {Bartlett}, {Bauer}, {Bauman}, {Baumont}, {Bechtol}, {Bechtol}, {Becker}, {Becla}, {Beldica}, {Bellavia}, {Bianco}, {Biswas}, {Blanc}, {Blazek}, {Blandford}, {Bloom}, {Bogart}, {Bond}, {Booth}, {Borgland}, {Borne}, {Bosch}, {Boutigny}, {Brackett}, {Bradshaw}, {Brandt}, {Brown}, {Bullock}, {Burchat}, {Burke}, {Cagnoli}, {Calabrese}, {Callahan}, {Callen}, {Carlin}, {Carlson}, {Chandrasekharan}, {Charles-Emerson}, {Chesley}, {Cheu}, {Chiang}, {Chiang}, {Chirino}, {Chow}, {Ciardi}, {Claver}, {Cohen-Tanugi}, {Cockrum}, {Coles}, {Connolly}, {Cook}, {Cooray}, {Covey}, {Cribbs}, {Cui}, {Cutri}, {Daly}, {Daniel}, {Daruich}, {Daubard}, {Daues}, {Dawson}, {Delgado}, {Dellapenna}, {de Peyster}, {de Val-Borro}, {Digel}, {Doherty}, {Dubois},
  {Dubois-Felsmann}, {Durech}, {Economou}, {Eifler}, {Eracleous}, {Emmons}, {Fausti Neto}, {Ferguson}, {Figueroa}, {Fisher-Levine}, {Focke}, {Foss}, {Frank}, {Freemon}, {Gangler}, {Gawiser}, {Geary}, {Gee}, {Geha}, {Gessner}, {Gibson}, {Gilmore}, {Glanzman}, {Glick}, {Goldina}, {Goldstein}, {Goodenow}, {Graham}, {Gressler}, {Gris}, {Guy}, {Guyonnet}, {Haller}, {Harris}, {Hascall}, {Haupt}, {Hernandez}, {Herrmann}, {Hileman}, {Hoblitt}, {Hodgson}, {Hogan}, {Howard}, {Huang}, {Huffer}, {Ingraham}, {Innes}, {Jacoby}, {Jain}, {Jammes}, {Jee}, {Jenness}, {Jernigan}, {Jevremovi{\'c}}, {Johns}, {Johnson}, {Johnson}, {Jones}, {Juramy-Gilles}, {Juri{\'c}}, {Kalirai}, {Kallivayalil}, {Kalmbach}, {Kantor}, {Karst}, {Kasliwal}, {Kelly}, {Kessler}, {Kinnison}, {Kirkby}, {Knox}, {Kotov}, {Krabbendam}, {Krughoff}, {Kub{\'a}nek}, {Kuczewski}, {Kulkarni}, {Ku}, {Kurita}, {Lage}, {Lambert}, {Lange}, {Langton}, {Le Guillou}, {Levine}, {Liang}, {Lim}, {Lintott}, {Long}, {Lopez}, {Lotz}, {Lupton}, {Lust}, {MacArthur}, {Mahabal},
  {Mandelbaum}, {Markiewicz}, {Marsh}, {Marshall}, {Marshall}, {May}, {McKercher}, {McQueen}, {Meyers}, {Migliore}, {Miller}, \& {Mills}}]{2019ApJ...873..111I}
{Ivezi{\'c}}, {\v{Z}}., {Kahn}, S.~M., {Tyson}, J.~A., {et~al.} 2019, \apj, 873, 111

\bibitem[{{Jiang} {et~al.}(2014){Jiang}, {Fan}, {Bian}, {McGreer}, {Strauss}, {Annis}, {Buck}, {Green}, {Hodge}, {Myers}, {Rafiee}, \& {Richards}}]{2014ApJS..213...12J}
{Jiang}, L., {Fan}, X., {Bian}, F., {et~al.} 2014, \apjs, 213, 12

\bibitem[{{Jo} {et~al.}(2018){Jo}, {Seon}, {Shinn}, {Yang}, {Lee}, \& {Min}}]{Jo2018}
{Jo}, Y.-S., {Seon}, K.-I., {Shinn}, J.-H., {et~al.} 2018, \apj, 862, 25

\bibitem[{{Johnston} {et~al.}(2008){Johnston}, {Bullock}, {Sharma}, {Font}, {Robertson}, \& {Leitner}}]{2008ApJ...689..936J}
{Johnston}, K.~V., {Bullock}, J.~S., {Sharma}, S., {et~al.} 2008, \apj, 689, 936

\bibitem[{{Ko} {et~al.}(2016){Ko}, {Chung}, {Hwang}, \& {Lee}}]{Ko+16}
{Ko}, J., {Chung}, H., {Hwang}, H.~S., \& {Lee}, J.~C. 2016, \apj, 820, 132

\bibitem[{{Ko} {et~al.}(2013){Ko}, {Hwang}, {Lee}, \& {Sohn}}]{Ko+13}
{Ko}, J., {Hwang}, H.~S., {Lee}, J.~C., \& {Sohn}, Y.-J. 2013, \apj, 767, 90

\bibitem[{{Ko} \& {Jee}(2018)}]{2018ApJ...862...95K}
{Ko}, J., \& {Jee}, M.~J. 2018, \apj, 862, 95

\bibitem[{{Kormendy} \& {Bahcall}(1974)}]{1974AJ.....79..671K}
{Kormendy}, J., \& {Bahcall}, J.~N. 1974, \aj, 79, 671

\bibitem[{{Laureijs} {et~al.}(2011){Laureijs}, {Amiaux}, {Arduini}, {Augu{\`e}res}, {Brinchmann}, {Cole}, {Cropper}, {Dabin}, {Duvet}, {Ealet}, {Garilli}, {Gondoin}, {Guzzo}, {Hoar}, {Hoekstra}, {Holmes}, {Kitching}, {Maciaszek}, {Mellier}, {Pasian}, {Percival}, {Rhodes}, {Saavedra Criado}, {Sauvage}, {Scaramella}, {Valenziano}, {Warren}, {Bender}, {Castander}, {Cimatti}, {Le F{\`e}vre}, {Kurki-Suonio}, {Levi}, {Lilje}, {Meylan}, {Nichol}, {Pedersen}, {Popa}, {Rebolo Lopez}, {Rix}, {Rottgering}, {Zeilinger}, {Grupp}, {Hudelot}, {Massey}, {Meneghetti}, {Miller}, {Paltani}, {Paulin-Henriksson}, {Pires}, {Saxton}, {Schrabback}, {Seidel}, {Walsh}, {Aghanim}, {Amendola}, {Bartlett}, {Baccigalupi}, {Beaulieu}, {Benabed}, {Cuby}, {Elbaz}, {Fosalba}, {Gavazzi}, {Helmi}, {Hook}, {Irwin}, {Kneib}, {Kunz}, {Mannucci}, {Moscardini}, {Tao}, {Teyssier}, {Weller}, {Zamorani}, {Zapatero Osorio}, {Boulade}, {Foumond}, {Di Giorgio}, {Guttridge}, {James}, {Kemp}, {Martignac}, {Spencer}, {Walton}, {Bl{\"u}mchen}, {Bonoli},
  {Bortoletto}, {Cerna}, {Corcione}, {Fabron}, {Jahnke}, {Ligori}, {Madrid}, {Martin}, {Morgante}, {Pamplona}, {Prieto}, {Riva}, {Toledo}, {Trifoglio}, {Zerbi}, {Abdalla}, {Douspis}, {Grenet}, {Borgani}, {Bouwens}, {Courbin}, {Delouis}, {Dubath}, {Fontana}, {Frailis}, {Grazian}, {Koppenh{\"o}fer}, {Mansutti}, {Melchior}, {Mignoli}, {Mohr}, {Neissner}, {Noddle}, {Poncet}, {Scodeggio}, {Serrano}, {Shane}, {Starck}, {Surace}, {Taylor}, {Verdoes-Kleijn}, {Vuerli}, {Williams}, {Zacchei}, {Altieri}, {Escudero Sanz}, {Kohley}, {Oosterbroek}, {Astier}, {Bacon}, {Bardelli}, {Baugh}, {Bellagamba}, {Benoist}, {Bianchi}, {Biviano}, {Branchini}, {Carbone}, {Cardone}, {Clements}, {Colombi}, {Conselice}, {Cresci}, {Deacon}, {Dunlop}, {Fedeli}, {Fontanot}, {Franzetti}, {Giocoli}, {Garcia-Bellido}, {Gow}, {Heavens}, {Hewett}, {Heymans}, {Holland}, {Huang}, {Ilbert}, {Joachimi}, {Jennins}, {Kerins}, {Kiessling}, {Kirk}, {Kotak}, {Krause}, {Lahav}, {van Leeuwen}, {Lesgourgues}, {Lombardi}, {Magliocchetti}, {Maguire},
  {Majerotto}, {Maoli}, {Marulli}, {Maurogordato}, {McCracken}, {McLure}, {Melchiorri}, {Merson}, {Moresco}, {Nonino}, {Norberg}, {Peacock}, {Pello}, {Penny}, {Pettorino}, {Di Porto}, {Pozzetti}, {Quercellini}, {Radovich}, {Rassat}, {Roche}, {Ronayette}, \& {Rossetti}}]{2011arXiv1110.3193L}
{Laureijs}, R., {Amiaux}, J., {Arduini}, S., {et~al.} 2011, arXiv e-prints, arXiv:1110.3193

\bibitem[{Layden {et~al.}(2025)Layden, Juneau, Pettersson, Lourie, Schneider, LaMarr, Angile, Farag, Luo, Ong, \& Fur{\'e}sz}]{Layden+25}
Layden, C., Juneau, J., Pettersson, G., {et~al.} 2025, Journal of Astronomical Telescopes, Instruments, and Systems, 11, 026003

\bibitem[{{Lee} {et~al.}(2022){Lee}, {Kimm}, {Blaizot}, {Katz}, {Lee}, {Sheen}, {Devriendt}, \& {Slyz}}]{lee22}
{Lee}, J., {Kimm}, T., {Blaizot}, J., {et~al.} 2022, \apj, 928, 144

\bibitem[{{Majewski} {et~al.}(2003){Majewski}, {Skrutskie}, {Weinberg}, \& {Ostheimer}}]{2003ApJ...599.1082M}
{Majewski}, S.~R., {Skrutskie}, M.~F., {Weinberg}, M.~D., \& {Ostheimer}, J.~C. 2003, \apj, 599, 1082

\bibitem[{{Malhan} {et~al.}(2022){Malhan}, {Ibata}, {Sharma}, {Famaey}, {Bellazzini}, {Carlberg}, {D'Souza}, {Yuan}, {Martin}, \& {Thomas}}]{Malhan2022}
{Malhan}, K., {Ibata}, R.~A., {Sharma}, S., {et~al.} 2022, \apj, 926, 107

\bibitem[{{Martin} {et~al.}(2022){Martin}, {Bazkiaei}, {Spavone}, {Iodice}, {Mihos}, {Montes}, {Benavides}, {Brough}, {Carlin}, {Collins}, {Duc}, {G{\'o}mez}, {Galaz}, {Hern{\'a}ndez-Toledo}, {Jackson}, {Kaviraj}, {Knapen}, {Mart{\'\i}nez-Lombilla}, {McGee}, {O'Ryan}, {Prole}, {Rich}, {Rom{\'a}n}, {Shah}, {Starkenburg}, {Watkins}, {Zaritsky}, {Pichon}, {Armus}, {Bianconi}, {Buitrago}, {Bus{\'a}}, {Davis}, {Demarco}, {Desmons}, {Garc{\'\i}a}, {Graham}, {Holwerda}, {Hon}, {Khalid}, {Klehammer}, {Klutse}, {Lazar}, {Nair}, {Noakes-Kettel}, {Rutkowski}, {Saha}, {Sahu}, {Sola}, {V{\'a}zquez-Mata}, {Vera-Casanova}, \& {Yoon}}]{2022MNRAS.513.1459M}
{Martin}, G., {Bazkiaei}, A.~E., {Spavone}, M., {et~al.} 2022, \mnras, 513, 1459

\bibitem[{{Mart{\'\i}nez-Delgado}(2019)}]{2019hsax.conf..146M}
{Mart{\'\i}nez-Delgado}, D. 2019, in Highlights on Spanish Astrophysics X, ed. B.~{Montesinos}, A.~{Asensio Ramos}, F.~{Buitrago}, R.~{Sch{\"o}del}, E.~{Villaver}, S.~{P{\'e}rez-Hoyos}, \& I.~{Ord{\'o}{\~n}ez-Etxeberria}, 146--154, \dodoi{10.48550/arXiv.1811.12286}

\bibitem[{{Mart{\'\i}nez-Delgado} {et~al.}(2023){Mart{\'\i}nez-Delgado}, {Cooper}, {Rom{\'a}n}, {Pillepich}, {Erkal}, {Pearson}, {Moustakas}, {Laporte}, {Laine}, {Akhlaghi}, {Lang}, {Makarov}, {Borlaff}, {Donatiello}, {Pearson}, {Mir{\'o}-Carretero}, {Cuillandre}, {Dom{\'\i}nguez}, {Roca-F{\`a}brega}, {Frenk}, {Schmidt}, {G{\'o}mez-Flechoso}, {Guzman}, {Libeskind}, {Dey}, {Weaver}, {Schlegel}, {Myers}, \& {Valdes}}]{2023A&A...671A.141M}
{Mart{\'\i}nez-Delgado}, D., {Cooper}, A.~P., {Rom{\'a}n}, J., {et~al.} 2023, \aap, 671, A141

\bibitem[{{Mart{\'\i}nez-Lombilla} {et~al.}(2023){Mart{\'\i}nez-Lombilla}, {Brough}, {Montes}, {Baena-Gall{\'e}}, {Akhlaghi}, {Infante-Sainz}, {Driver}, {Holwerda}, {Pimbblet}, \& {Robotham}}]{2023MNRAS.518.1195M}
{Mart{\'\i}nez-Lombilla}, C., {Brough}, S., {Montes}, M., {et~al.} 2023, \mnras, 518, 1195

\bibitem[{{Merritt} {et~al.}(2014){Merritt}, {van Dokkum}, \& {Abraham}}]{2014ApJ...787L..37M}
{Merritt}, A., {van Dokkum}, P., \& {Abraham}, R. 2014, \apjl, 787, L37

\bibitem[{{Merritt} {et~al.}(2016){Merritt}, {van Dokkum}, {Abraham}, \& {Zhang}}]{2016ApJ...830...62M}
{Merritt}, A., {van Dokkum}, P., {Abraham}, R., \& {Zhang}, J. 2016, \apj, 830, 62

\bibitem[{{Mihos}(2019)}]{2019arXiv190909456M}
{Mihos}, J.~C. 2019, arXiv e-prints, arXiv:1909.09456

\bibitem[{{Mihos} {et~al.}(2017){Mihos}, {Harding}, {Feldmeier}, {Rudick}, {Janowiecki}, {Morrison}, {Slater}, \& {Watkins}}]{2017ApJ...834...16M}
{Mihos}, J.~C., {Harding}, P., {Feldmeier}, J.~J., {et~al.} 2017, \apj, 834, 16

\bibitem[{{Park} {et~al.}(2020){Park}, {Chang}, {Lim}, {Lee}, {Ahn}, {Kim}, {Kim}, {Hammar}, {Jeong}, {Kim}, {Lee}, {Kim}, \& {Pak}}]{2020PASP..132d4504P}
{Park}, W., {Chang}, S., {Lim}, J.~H., {et~al.} 2020, \pasp, 132, 044504

\bibitem[{{Parker} {et~al.}(2020){Parker}, {Sechrest}, {Vestrand}, {Wozniak}, \& {Palmer}}]{2020SPIE11447E..A2P}
{Parker}, L.~P., {Sechrest}, Y.~H., {Vestrand}, W.~T., {Wozniak}, P., \& {Palmer}, D. 2020, in Society of Photo-Optical Instrumentation Engineers (SPIE) Conference Series, Vol. 11447, Ground-based and Airborne Instrumentation for Astronomy VIII, ed. C.~J. {Evans}, J.~J. {Bryant}, \& K.~{Motohara}, 11447A2, \dodoi{10.1117/12.2576330}

\bibitem[{{Perlmutter} {et~al.}(1999){Perlmutter}, {Aldering}, {Goldhaber}, {Knop}, {Nugent}, {Castro}, {Deustua}, {Fabbro}, {Goobar}, {Groom}, {Hook}, {Kim}, {Kim}, {Lee}, {Nunes}, {Pain}, {Pennypacker}, {Quimby}, {Lidman}, {Ellis}, {Irwin}, {McMahon}, {Ruiz-Lapuente}, {Walton}, {Schaefer}, {Boyle}, {Filippenko}, {Matheson}, {Fruchter}, {Panagia}, {Newberg}, {Couch}, \& {Project}}]{Perlmutter1999}
{Perlmutter}, S., {Aldering}, G., {Goldhaber}, G., {et~al.} 1999, \apj, 517, 565

\bibitem[{{Pillepich} {et~al.}(2014){Pillepich}, {Vogelsberger}, {Deason}, {Rodriguez-Gomez}, {Genel}, {Nelson}, {Torrey}, {Sales}, {Marinacci}, {Springel}, {Sijacki}, \& {Hernquist}}]{2014MNRAS.444..237P}
{Pillepich}, A., {Vogelsberger}, M., {Deason}, A., {et~al.} 2014, \mnras, 444, 237

\bibitem[{{Purcell} {et~al.}(2007){Purcell}, {Bullock}, \& {Zentner}}]{2007ApJ...666...20P}
{Purcell}, C.~W., {Bullock}, J.~S., \& {Zentner}, A.~R. 2007, \apj, 666, 20

\bibitem[{{Rich} {et~al.}(2019){Rich}, {Mosenkov}, {Lee-Saunders}, {Koch}, {Kormendy}, {Kennefick}, {Brosch}, {Sales}, {Bullock}, {Burkert}, {Collins}, {Cooper}, {Fusco}, {Reitzel}, {Thilker}, {Milewski}, {Elias}, {Saade}, \& {De Groot}}]{2019MNRAS.490.1539R}
{Rich}, R.~M., {Mosenkov}, A., {Lee-Saunders}, H., {et~al.} 2019, \mnras, 490, 1539

\bibitem[{{Riess} {et~al.}(1998){Riess}, {Filippenko}, {Challis}, {Clocchiatti}, {Diercks}, {Garnavich}, {Gilliland}, {Hogan}, {Jha}, {Kirshner}, {Leibundgut}, {Phillips}, {Reiss}, {Schmidt}, {Schommer}, {Smith}, {Spyromilio}, {Stubbs}, {Suntzeff}, \& {Tonry}}]{Riess1998}
{Riess}, A.~G., {Filippenko}, A.~V., {Challis}, P., {et~al.} 1998, \aj, 116, 1009

\bibitem[{{Rimoldini} {et~al.}(2023){Rimoldini}, {Holl}, {Gavras}, {Audard}, {De Ridder}, {Mowlavi}, {Nienartowicz}, {Jevardat de Fombelle}, {Lecoeur-Ta{\"\i}bi}, {Karbevska}, {Evans}, {{\'A}brah{\'a}m}, {Carnerero}, {Clementini}, {Distefano}, {Garofalo}, {Garc{\'\i}a-Lario}, {Gomel}, {Klioner}, {Kruszy{\'n}ska}, {Lanzafame}, {Lebzelter}, {Marton}, {Mazeh}, {Molinaro}, {Panahi}, {Raiteri}, {Ripepi}, {Szabados}, {Teyssier}, {Trabucchi}, {Wyrzykowski}, {Zucker}, \& {Eyer}}]{Rimold2023}
{Rimoldini}, L., {Holl}, B., {Gavras}, P., {et~al.} 2023, \aap, 674, A14

\bibitem[{{Rom{\'a}n} {et~al.}(2020){Rom{\'a}n}, {Trujillo}, \& {Montes}}]{2020A&A...644A..42R}
{Rom{\'a}n}, J., {Trujillo}, I., \& {Montes}, M. 2020, \aap, 644, A42

\bibitem[{{Sales} {et~al.}(2022){Sales}, {Wetzel}, \& {Fattahi}}]{2022NatAs...6..897S}
{Sales}, L.~V., {Wetzel}, A., \& {Fattahi}, A. 2022, Nature Astronomy, 6, 897

\bibitem[{{Sandage} \& {Binggeli}(1984)}]{1984AJ.....89..919S}
{Sandage}, A., \& {Binggeli}, B. 1984, \aj, 89, 919

\bibitem[{{Shen} {et~al.}(2024){Shen}, {Bowman}, {van Dokkum}, {Abraham}, {Pasha}, {Keim}, {Liu}, {Lokhorst}, {Janssens}, \& {Chen}}]{2024ApJ...976...75S}
{Shen}, Z., {Bowman}, W.~P., {van Dokkum}, P., {et~al.} 2024, \apj, 976, 75

\bibitem[{{Shipp} {et~al.}(2023){Shipp}, {Panithanpaisal}, {Necib}, {Sanderson}, {Erkal}, {Li}, {Santistevan}, {Wetzel}, {Cullinane}, {Ji}, {Koposov}, {Kuehn}, {Lewis}, {Pace}, {Zucker}, {Bland-Hawthorn}, {Cunningham}, {Kim}, {Lilleengen}, {Moreno}, {Sharma}, {S Collaboration}, \& {FIRE Collaboration}}]{Shipp2023}
{Shipp}, N., {Panithanpaisal}, N., {Necib}, L., {et~al.} 2023, \apj, 949, 44

\bibitem[{{Sola} {et~al.}(2022){Sola}, {Duc}, {Richards}, {Paiement}, {Urbano}, {Klehammer}, {B{\'\i}lek}, {Cuillandre}, {Gwyn}, \& {McConnachie}}]{2022A&A...662A.124S}
{Sola}, E., {Duc}, P.-A., {Richards}, F., {et~al.} 2022, \aap, 662, A124

\bibitem[{{Trujillo} \& {Fliri}(2016)}]{2016ApJ...823..123T}
{Trujillo}, I., \& {Fliri}, J. 2016, \apj, 823, 123

\bibitem[{{van Dokkum} {et~al.}(2015){van Dokkum}, {Abraham}, {Merritt}, {Zhang}, {Geha}, \& {Conroy}}]{2015ApJ...798L..45V}
{van Dokkum}, P.~G., {Abraham}, R., {Merritt}, A., {et~al.} 2015, \apjl, 798, L45

\bibitem[{{Yagi} {et~al.}(2017){Yagi}, {Yoshida}, {Gavazzi}, {Komiyama}, {Kashikawa}, \& {Okamura}}]{yagi17}
{Yagi}, M., {Yoshida}, M., {Gavazzi}, G., {et~al.} 2017, \apj, 839, 65

\bibitem[{{Yagi} {et~al.}(2010){Yagi}, {Yoshida}, {Komiyama}, {Kashikawa}, {Furusawa}, {Okamura}, {Graham}, {Miller}, {Carter}, {Mobasher}, \& {Jogee}}]{yagi10}
{Yagi}, M., {Yoshida}, M., {Komiyama}, Y., {et~al.} 2010, \aj, 140, 1814

\bibitem[{{Yi} {et~al.}(2005){Yi}, {Yoon}, {Kaviraj}, {Deharveng}, {Rich}, {Salim}, {Boselli}, {Lee}, {Ree}, {Sohn}, {Rey}, {Lee}, {Rhee}, {Bianchi}, {Byun}, {Donas}, {Friedman}, {Heckman}, {Jelinsky}, {Madore}, {Malina}, {Martin}, {Milliard}, {Morrissey}, {Neff}, {Schiminovich}, {Siegmund}, {Small}, {Szalay}, {Jee}, {Kim}, {Barlow}, {Forster}, {Welsh}, \& {Wyder}}]{Yi+05}
{Yi}, S.~K., {Yoon}, S.~J., {Kaviraj}, S., {et~al.} 2005, \apjl, 619, L111

\bibitem[{{Yoo} {et~al.}(2025){Yoo}, {Shin}, {Hwang}, {Sabiu}, {Kim}, {Ko}, \& {Lee}}]{Yoo2025}
{Yoo}, J., {Shin}, J., {Hwang}, H.~S., {et~al.} 2025, \apj, 988, 229

\bibitem[{{Yoo} {et~al.}(2022){Yoo}, {Ko}, {Sabiu}, {Shin}, {Chun}, {Hwang}, {Kim}, {Jee}, {Kim}, \& {Smith}}]{2022ApJS..261...28Y}
{Yoo}, J., {Ko}, J., {Sabiu}, C.~G., {et~al.} 2022, \apjs, 261, 28

\bibitem[{{Yoo} {et~al.}(2024){Yoo}, {Park}, {Sabiu}, {Singh}, {Ko}, {Lee}, {Pichon}, {Jee}, {Gibson}, {Snaith}, {Kim}, {Shin}, {Kim}, \& {Kim}}]{Yoo2024}
{Yoo}, J., {Park}, C., {Sabiu}, C.~G., {et~al.} 2024, \apj, 965, 145

\bibitem[{{York} {et~al.}(2000){York}, {Adelman}, {Anderson}, {Anderson}, {Annis}, {Bahcall}, {Bakken}, {Barkhouser}, {Bastian}, {Berman}, {Boroski}, {Bracker}, {Briegel}, {Briggs}, {Brinkmann}, {Brunner}, {Burles}, {Carey}, {Carr}, {Castander}, {Chen}, {Colestock}, {Connolly}, {Crocker}, {Csabai}, {Czarapata}, {Davis}, {Doi}, {Dombeck}, {Eisenstein}, {Ellman}, {Elms}, {Evans}, {Fan}, {Federwitz}, {Fiscelli}, {Friedman}, {Frieman}, {Fukugita}, {Gillespie}, {Gunn}, {Gurbani}, {de Haas}, {Haldeman}, {Harris}, {Hayes}, {Heckman}, {Hennessy}, {Hindsley}, {Holm}, {Holmgren}, {Huang}, {Hull}, {Husby}, {Ichikawa}, {Ichikawa}, {Ivezi{\'c}}, {Kent}, {Kim}, {Kinney}, {Klaene}, {Kleinman}, {Kleinman}, {Knapp}, {Korienek}, {Kron}, {Kunszt}, {Lamb}, {Lee}, {Leger}, {Limmongkol}, {Lindenmeyer}, {Long}, {Loomis}, {Loveday}, {Lucinio}, {Lupton}, {MacKinnon}, {Mannery}, {Mantsch}, {Margon}, {McGehee}, {McKay}, {Meiksin}, {Merelli}, {Monet}, {Munn}, {Narayanan}, {Nash}, {Neilsen}, {Neswold}, {Newberg}, {Nichol}, {Nicinski},
  {Nonino}, {Okada}, {Okamura}, {Ostriker}, {Owen}, {Pauls}, {Peoples}, {Peterson}, {Petravick}, {Pier}, {Pope}, {Pordes}, {Prosapio}, {Rechenmacher}, {Quinn}, {Richards}, {Richmond}, {Rivetta}, {Rockosi}, {Ruthmansdorfer}, {Sandford}, {Schlegel}, {Schneider}, {Sekiguchi}, {Sergey}, {Shimasaku}, {Siegmund}, {Smee}, {Smith}, {Snedden}, {Stone}, {Stoughton}, {Strauss}, {Stubbs}, {SubbaRao}, {Szalay}, {Szapudi}, {Szokoly}, {Thakar}, {Tremonti}, {Tucker}, {Uomoto}, {Vanden Berk}, {Vogeley}, {Waddell}, {Wang}, {Watanabe}, {Weinberg}, {Yanny}, {Yasuda}, \& {SDSS Collaboration}}]{2000AJ....120.1579Y}
{York}, D.~G., {Adelman}, J., {Anderson}, Jr., J.~E., {et~al.} 2000, \aj, 120, 1579

\end{thebibliography}






\end{document}